\documentclass{MITcsail}

\usepackage{paracol}  
\usepackage{graphicx} 
\usepackage{ragged2e} 
\usepackage[spanish, english]{babel} 
\usepackage{setspace}

\usepackage{natbib}
\setcitestyle{authoryear,open={(},close={)}} 
\usepackage{multirow}

\title{Inteligencia Artificial para la conservación y uso sostenible de la biodiversidad, una visión desde Colombia \\ \normalfont{Artificial Intelligence for conservation and \\sustainable use of biodiversity, a view from Colombia}}

\author{Juan Sebastián Cañas$^{1,2}$\footnote{Corresponding author(s). E-mail(s): juan.canas@ucl.ac.uk; julloa@humboldt.org.co.}, Camila Parra-Guevara$^{1,3}$, Manuela Montoya-Castrillón$^{1}$, Julieta M Ramírez-Mejía$^{1}$, Gabriel-Alejandro Perilla$^{1}$, Esteban Marentes$^{1}$, Nerieth Leuro$^{1,3}$, Jose Vladimir Sandoval-Sierra$^{1}$, Sindy Martinez-Callejas$^{1}$, Angélica Díaz$^{1}$, Mario Murcia$^{1}$, Elkin A. Noguera-Urbano$^{1}$, Jose Manuel Ochoa-Quintero$^{1,4}$, Susana Rodríguez Buriticá$^{1}$, Juan Sebastián Ulloa$^{1}$\footnotemark[1]%
\vspace{1em} 
 \\ \normalfont{\small $^{1}$ \textit{Instituto de Investigación de Recursos Biológicos Alexander von Humboldt, Calle 72 \#12-65 Piso 7, Bogotá, Colombia.}} 
 \\ \normalfont{\small $^{2}$ \textit{Centre for Biodiversity and Environment Research (CBER),\color{white} ..\color{black} Department of Genetics, Evolution and Environment, \color{white} . . . \color{black} University College London, London, WC1E 6BT, United Kingdom.}} 
 \\ \normalfont{\small $^{3}$ \textit{Sistema de Información sobre Biodiversidad de Colombia (SiB Colombia), Calle 72 \#12-65 Piso 7, Bogotá, Colombia.}}
 \\ \normalfont{\small $^{4}$ \textit{Instituto de Biociências, Universidade Federal de Mato Grosso do Sul, Campo Grande, Brasil.}}\\
}

\footnotelayout{m}

\begin{document}

\maketitle
\columnratio{0.521}
\begin{paracol}{2}

\justifying

\textbf{Resumen:} El auge de la inteligencia artificial (IA) y el agravamiento de la crisis de la biodiversidad ha dado como resultado un área de investigación en la que se desarrollan métodos computacionales basados en IA que actúan como aliados en la conservación, y el aprovechamiento y uso sostenible de los recursos naturales. Aunque importantes lineamientos generales se han establecido a nivel global sobre oportunidades y desafíos que esta investigación interdisciplinaria ofrece, es importante generar reflexiones locales desde los contextos y realidades específicas de cada región. Por ello, este documento tiene como objetivo anali\-zar los alcances de esta área de investigación desde una perspectiva centrada en Colombia y el Neotrópico. En este documento resumimos las principales experiencias y debates que se llevaron a cabo en el Instituto Humboldt en Colombia entre 2023 y 2024. Para ejemplificar la multiplicidad de oportunidades prometedoras, se presentan usos actuales, como la identificación automática de especies a partir de imágenes y grabaciones, el modelado de especies, y la bioprospección in silico, entre otros. A partir de las experiencias descritas anteriormente, destacamos limitaciones, desafíos y oportunidades para implementar exitosamente la IA en labores de conservación y manejo sostenible de recursos biológicos en el Neotrópico. El resultado pretende ser una guía para investigadores, tomadores de decisiones y gestores de la biodiversidad, que facilite la comprensión de cómo la IA  puede integrarse eficazmente en estrategias de conservación y uso sostenible. Asimismo, busca abrir un espacio para el diálogo sobre el desarrollo de políticas que promuevan la adopción responsable y ética de la IA en contextos locales, de manera que se aprovechen sus beneficios sin comprometer la biodiversidad ni los valores culturales y ecosistémicos propios de Colombia y el Neotrópico.

\textbf{Palabras Clave:} Aprendizaje Automático, Monitoreo de la Biodiversidad, Toma de Decisiones Ambientales, Ciencia, Tecnología y Sociedad, Sistemas de IA.

\switchcolumn 

\justifying

\textbf{Abstract:} The rise of artificial intelligence (AI) and the aggravating biodiversity crisis have resulted in a research area where AI-based computational methods are being developed to act as allies in conservation, and the sustainable use and management of natural resources. While important general guidelines have been established globally regarding the opportunities and challenges that this interdisciplinary research offers, it is essential to generate local reflections from the specific contexts and realities of each region. Hence, this document aims to analyze the scope of this research area from a perspective focused on Colombia and the Neotropics. In this paper, we summarize the main experiences and debates that took place at the Humboldt Institute between 2023 and 2024 in Colombia. To illustrate the variety of promising opportunities, we present current uses such as automatic species identification from images and recordings, species modeling, and in silico bioprospecting, among others. From the experiences described above, we highlight limitations, challenges, and opportunities for in order to successfully implementate AI in conservation efforts and sustainable management of biological resources in the Neotropics. The result aims to be a guide for researchers, decision makers, and biodiversity managers, facilitating the understanding of how artificial intelligence can be effectively integrated into conservation and sustainable use strategies. Furthermore, it also seeks to open a space for dialogue on the development of policies that promote the responsible and ethical adoption of AI in local contexts, ensuring that its benefits are harnessed without compromising biodiversity or the cultural and ecosystemic values inherent in Colombia and the Neotropics.   


\end{paracol}


\begin{paracol}{2} 

\section*{Introducción}

\justifying

Durante la última década, la Inteligencia Artificial (IA) ha impulsado avances significativos en sectores como la medicina, el comercio, la educación, e incluso las artes. Desde diagnósticos médicos más precisos y rápidos, hasta sistemas de recomendación personalizados y análisis de grandes volúmenes de datos, la IA ha transformado la forma en que interactuamos con el mundo y tomamos decisiones. Con este precedente, es crucial indagar cómo esta tecnología puede adaptarse para enfrentar grandes retos globales: conservar la biodiversidad, promover el uso sostenible de sus componentes y asegurar la distribución justa y equitativa de los beneficios derivados de los recursos naturales.

La IA se puede definir como la capacidad que tiene una máquina para realizar razonamientos semánticos de alto nivel. Esto es, la capacidad de un sistema de interpretar información existente, y proporcionar inferencias novedosas 
\citep{russell2016}. A partir del aumento de la capacidad de cómputo, la disponibilidad masiva de datos y los avances en algoritmos de aprendizaje profundo \citep{lecun2015} (ver Glosario de términos en el material suplementario \ref{sec:supplementary} para los términos más usados durante el texto), la IA ha mostrado un crecimiento exponencial en la industria, impulsando la innovación en sectores como la salud, la manufactura, las finanzas, el transporte y la comunicación. Estos avances demuestran el enorme potencial que tiene para abordar problemas complejos y mejorar la eficiencia en múltiples áreas, y explica su adopción en la cultura de masas. Cada día es más frecuente encontrarse con la IA como un tema recurrente en el paisaje urbano y digital. Actualmente, se promociona como una herramienta innovadora que enriquece y transforma cualquier ámbito en el que se aplica. En consecuencia, existen grandes expectativas y promesas sobre lo que es y lo que puede hacer, aunque el entendimiento de cómo funciona, cómo usarla y sus implicaciones sigue siendo, para muchos, bastante limitado \\ \citep{crawford2021}.

Más allá de un objeto que viene a reemplazar labores humanas, la IA puede ser diseñada como un objeto que potencia nuestros esfuerzos en búsqueda de soluciones para enfrentar desafíos globales y locales. La pérdida de biodiversidad y el cambio climático amenazan la estabilidad de nuestra sociedad, especialmente en países megadiversos como Colombia \citep{richardson2023}. Ante esta realidad, se han comenzado a explorar soluciones computacionales que aborden estos retos y herramientas que puedan apoyar estrategias de mitigación. Algunos ejemplos son la identificación de tendencias en datos medioambientales y su monitoreo, y el desarrollo de sistemas de alerta temprana para eventos naturales extremos donde la IA puede llegar a jugar un papel crucial en la protección y conservación de nuestro entorno. Sin embargo, aún es necesario un esfuerzo concertado para determinar la mejor manera de aplicar estas herramientas. La IA, como cualquier tecnología, tiene límites que son cruciales comprender, ya que desconocerlos puede resultar en riesgos considerables. Por ejemplo, los novedosos modelos de aprendizaje profundo tienen altas demandas energéticas y de consumo de agua para su funcionamiento \citep{bommasani2022,bender2021, varoquaux2024}. 

Asimismo, los modelos de aprendizaje de má\-quina están basados en aprender de ejemplos, estos ejemplos no son objetos aislados del mundo, sino que se tienen un proceso de generación embebido dentro de  un contexto social, por lo tanto, los modelos basados en datos naturalmente replican los distintos tipos de sesgos provenientes de estos datos \citep{barocas2023}. Adicionalmente, la IA como herramienta aplicada a sistemas socio-ecológicos tiene como resultado, sistemas socio-tecno-ecológicos \textit{-sistemas de IA-} los cuales tienen relaciones complejas e impactos a diferentes niveles y diferentes actores, creando nuevos retos \citep{rohde2024}. Debido a esto, es fundamental conocer cómo se puede adaptar a contextos específicos, como los países del Sur Global ubicados en el Neotrópico en los que converge una notoria diversidad biológica y humana, pero también dramáticos conflictos socioambientales y dramáticas dificultades económicas.

\switchcolumn

\section*{Introduction}
Over the past decade, Artificial Intelligence (AI) has driven significant advancements in sectors such as medicine, commerce, education, and even the arts. Ranging from more accurate and faster medical diagnoses to personalized recommendation systems and large-scale data analysis, AI has transformed the way we interact with the world and make decisions. All things considered, it is crucial to investigate how this technology can be adapted to address major global challenges: conserving biodiversity, promoting the sustainable use of its components, and ensuring the fair and equitable distribution of benefits derived from natural resources.   

AI can be defined as the ability of a machine to perform high-level semantic reasoning. That is, the ability of a system to interpret existing information and provide novel inferences \citep{russell2016}. With the increase in computing power, the massive availability of data, and advances in deep learning algorithms \citep{lecun2015} (refer to the Glossary of terms in the supplementary material \ref{sec:supplementary} for the most commonly used concepts throughout the paper), AI has experienced exponential growth in the industry, driving innovation in sectors such as health, manufacturing, finance, transportation, and communication. These advancements demonstrate its enormous potential to tackle complex problems and improve efficiency in multiple areas, which explains its widespread adoption in mass culture. It is increasingly common to see AI as a recurring theme in both urban and digital landscapes. It is currently promoted as an innovative tool that enriches and transforms any field in which it is applied. As a result, there are great expectations and promises about what AI is and what it can do, though the understanding of how it works, how to use it, and its implications remains, for many, quite limited \citep{crawford2021}.      

Beyond being a tool that replaces human labor, AI can be designed as a tool that enhances our efforts to find solutions to global and local challenges. The loss of biodiversity and climate  change threaten the stability of our society, especially in megadiverse countries like Colombia \citep{richardson2023}. In the light of this reality, computational solutions are being explored to address these challenges and tools that can support mitigation strategies. Some examples include identifying trends in environmental data and monitoring them, as well as developing early warning systems for extreme natural events where AI can play a crucial role in protecting and conserving our environment. However, there is still a need for a concerted effort to determine the best way to apply these tools. Artificial intelligence, like any technology, has limitations that are crucial to understand, as ignoring them can result in significant risks. For example, the novel deep learning models have high energy and water consumption demands for their operation 
\citep{bommasani2022,bender2021, varoquaux2024}. 
\\
\\

Moreover, machine learning models are based on learning from examples, which are not isolated objects from the world but are embedded in a generation process within a social context. Therefore, data-driven models naturally replicate the various biases originating from these data \citep{barocas2023}. Additionally, AI as a tool applied to socio-ecological systems results in socio-techno-ecological systems \textit{-AI systems-} which have complex relationships and impacts at different levels and different actors, creating new challenges \citep{rohde2024}. Because of this, it is essential to understand how AI can be adapted to specific contexts, such as the countries of the Global South located in the Neotropics, where notable biological and human diversity converge, as well as dramatic socio-environmental conflicts and economic difficulties.
\end{paracol}

\begin{figure}[t]
    \centering
    \includegraphics[width=1\textwidth]{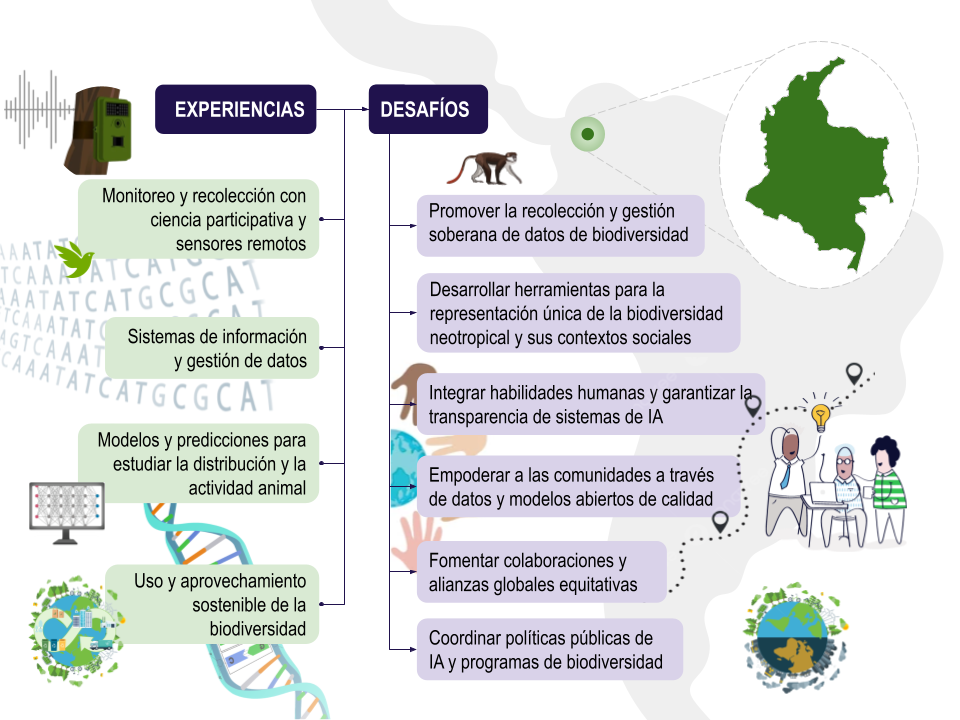}
    \caption{Experiencias y desafíos encontrados en el uso de la Inteligencia Artificial para la conservación y el uso sostenible de la biodiversidad en Colombia y el Neotrópico. \textit{Figure \ref{fig:main_figure} is translated in the Supplementary material as Figure \ref{fig:main_figure_eng}.}}
    \label{fig:main_figure}
\end{figure}

\begin{figure}[h]
    \centering
    \includegraphics[width=1\textwidth]{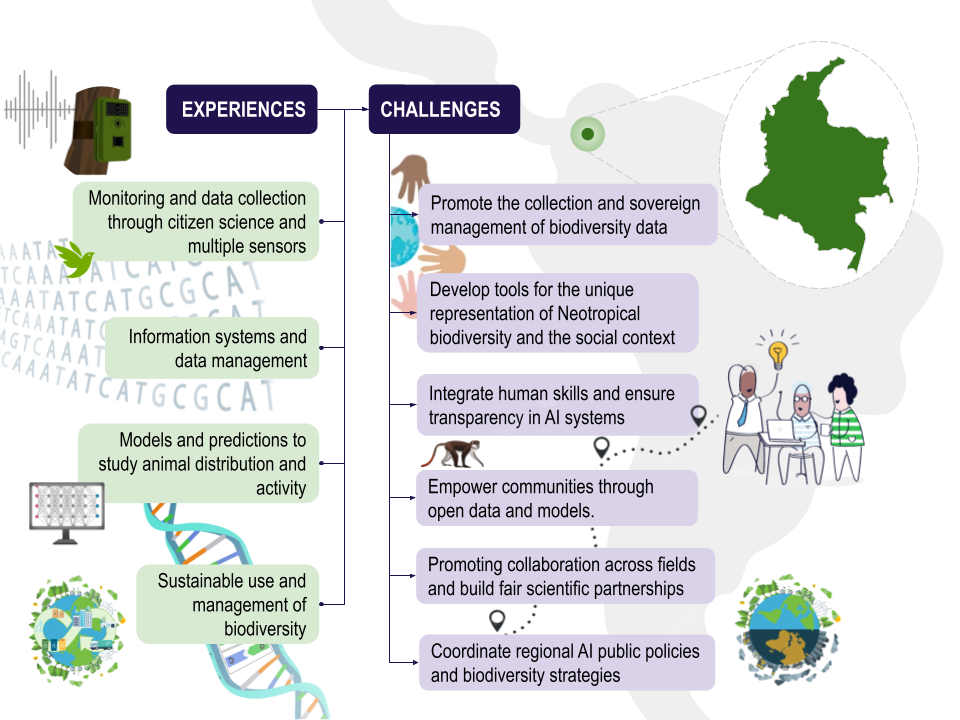}
    \caption{ Experiences and challenges encountered in the use of Artificial Intelligence for the conservation and sustainable use of biodiversity in the Neotropics. Translation of Figure \ref{fig:main_figure}.}
    \label{fig:main_figure_eng}
\end{figure}

\begin{paracol}{2}

La convergencia entre el desarrollo de tecnologías de IA y la pérdida de la biodiversidad, ha consolidado un área de investigación interdisciplinar donde las comunidades de ciencias de la computación, junto con las ciencias biológicas y sociales, crean sinergia \citep{beery2021scaling, tuia2022, rolnick2023, tambe2022}. Allí no sólo emergen soluciones novedosas que tienen un impacto en la sociedad y en la conservación, sino que plantea nuevos desafíos conceptuales y metodológicos dentro de la computación. Es decir, la necesidad de modelar sistemas ecológicos complejos, procesar datos ambientales ruidosos o incompletos y optimizar estrategias de conservación lleva a desarrollar nuevas teorías y enfoques en IA, aprendizaje automático y análisis de datos. A raíz del crecimiento de esta área, se han publicado reflexiones, limitaciones y potenciales \citep{gpai2022, xu2023, reynolds2024}, lo que ha servido para trazar diversas líneas de investigación e incluso delinear políticas públicas. No obstante, estas reflexiones provienen desde una visión del Norte Global. Este documento busca diversificar el debate, promoviendo perspectivas alternativas y contrarrestando la monocultura científica \citep{messeri2024artificial}. Para ello, exploramos el potencial de la IA en la conservación de la biodiversidad desde una perspectiva del y para Colombia y el Neotrópico. Nuestro análisis se nutre de la colaboración entre científicas de la computación y profesionales de la biodiversidad, considerando las diferencias contextuales y necesidades de esta región. 

El objetivo de este trabajo es realizar un diagnóstico del uso actual de la IA en la conservación de la biodiversidad en Colombia, con el fin de detectar desafíos y direcciones futuras para el Neotrópico. Asimismo, buscamos identificar posibles aplicaciones de la IA que necesitan mayores esfuerzos, y enfatizar en limitaciones y riesgos asociados a esta tecnología (ver Figura \ref{fig:main_figure}). Para esto, se realizó un taller en el Instituto de Investigación de Recursos Biológicos Alexander von Humboldt (Colombia), en el que se discutieron diversos proyectos y posibles líneas de trabajo en la intersección entre IA y conservación. Este trabajo no pretende identificar todas las aplicaciones existentes en el país, sino ser el primer paso para entender las posibles aplicaciones de la IA en el contexto Colombiano. Por ello, este trabajo está dirigido a la diversa comunidad de actores que se dedica a la conservación de la biodiversidad pero también a la comunidad de ciencias de la computación que están impulsando el uso y la aplicación de nuevas tecnologías. Esperamos que este documento sea una base para definir una estrategia de trabajo coordinada en torno a la IA para la conservación de la biodiversidad en los países del Neotrópico.

\switchcolumn 

The convergence between AI technology development and biodiversity loss has established a interdisciplinary research area where computer science communities, alongside biological and social sciences, create synergy \citep{beery2021scaling, rolnick2023,tambe2022, tuia2022}. Here, not only do novel solutions emerge that impact society and conservation, but it also presents new conceptual and methodological challenges within computation. That is, the need to model complex ecological systems, process noisy or incomplete environmental data, and optimize conservation strategies leads to the development of new theories and approaches in AI, machine learning, and data analysis. Due to the growth of this area, reflections, limitations, and potentials have been published \citep{gpai2022, xu2023, reynolds2024}, which have helped outline various research directions and even shape public policies. However, these reflections come from a Global North perspective. We seek to diversify the debate by promoting alternative perspectives opposed to scientific monoculture \citep{messeri2024artificial}. 
 Therefore, this document aims to discuss the scope of AI for biodiversity conservation from a perspective centered on and for Colombia and the Neotropics. Our analysis is informed by collaboration between computer scientists and biodiversity professionals, considering the contextual differences and needs of this region.   

This work aims to diagnose the current use of AI for biodiversity conservation in Colombia, in order to identify challenges and future directions for the Neotropics. Additionally, we seek to identify potential applications of AI that require greater efforts and emphasize the limitations and risks associated with this technology (check Figure \ref{fig:main_figure_eng}). To this end, a workshop was held at the Alexander von Humboldt Biological Resources Research Institute (Colombia), where various projects and potential lines of work at the intersection of AI and conservation were discussed. This work does not aim to identify all existing applications in the country but to be the first step in understanding the possible applications of AI in the Colombian context. Therefore, this paper is directed at the diverse community of actors involved in biodiversity conservation, as well as the computer science community driving the use and application of new technologies. We hope that this document will serve as a foundation to define a coordinated work strategy around AI for biodiversity conservation in Neotropical countries.

\end{paracol}

\begin{paracol}{2}

\section*{Experiencias y aplicaciones}

En esta sección se presentan casos de estudio y aplicaciones de la inteligencia artificial en el ámbito de la conservación y el uso sostenible de la biodiversidad. Lo anterior, por medio de ejemplos prácticos desde diversas perspectivas que incluyen el monitoreo de la biodiversidad, la modelización de registros biológicos, la bioeconomía, el análisis de datos genéticos, entre otros. Cada caso ofrece un contexto específico, mostrando cómo la IA aporta soluciones innovadoras y cuáles lecciones aprendidas se pueden destacar de su implementación. Es así como a partir de estos estudios, una revisión bibliográfica y discusiones del equipo, se identificaron los principales desafíos y direcciones futuras a considerar.

\subsection*{1. Fototrampeo e IA al servicio de la conservación de la naturaleza}

\subsubsection*{Contexto}

Las especies de mamíferos y aves de hábitos crípti\-cos y con tamaños poblacionales pequeños fueron difícilmente estudiadas hasta que las cámaras trampa permitieron reducir el impacto de la presencia humana para la obtención de registros fotográficos (de especies crípticas) que tengan asociados metadatos espaciales y temporales, entre otros. Es así como los datos obtenidos se han venido incrementando exponencialmente, en la medida que ha avanzado el desarrollo tecnológico de esta herramienta, la cual actualmente permite la recolección de imágenes  al menos por un mes, sin requerir de la presencia humana \citep{oliver2023b}. Esto ha reducido sus costos, por lo que es posible contar con un mayor número de equipos por estudio. Ante este panorama de datos se requiere de una gestión eficiente de los mismos, buscando optimizar cada uno de los procesos que se requieren para realizar análisis y obtener conclusiones que aporten a la toma de decisiones para la conservación de la naturaleza. 

\subsubsection*{Solución propuesta}
La interdisciplinariedad y el trabajo colaborativo hacen parte de la estrategia de investigación que ha permitido avanzar acertadamente en la optimización de procesos. El Instituto Humboldt ha impulsado colaboraciones para la creación de una aplicación de escritorio llamada NAIRA, llegando a su tercera versión en colaboración con la Universidad de Antioquia. NAIRA III es un software diseñado para facilitar el post-procesamiento de las imágenes obtenidas con cámaras trampa y ofrece al usuario la posibilidad de clasificar las imágenes identificando los animales presentes en cada fotografía y separando aquellas imágenes en las cuales no hay presencia de fauna \citep{giraldo2019, pulido2018}. Esta versión identifica automáticamente las fotografías con presencia de aves y mamíferos, y dentro de este último grupo a 20 géneros taxonómicos; extrae automáticamente la información de las fotografías como fase lunar, temperatura, coordenadas del GPS, fecha y hora de captura, y la organiza en datos tabulares siguiendo el estándar Darwin Core. Adicionalmente, calcula automáticamente la tabla de registros independientes junto con las matrices de presencia - ausencia y de conteos por fechas, fases lunares y temperatura. Otra función del programa es que permite al usuario agregar manualmente información que considere importante para el análisis de cada una de las sesiones de fototrampeo y realizar cambios en la clasificación obtenida de manera automática.
La IA, y los sistemas de organización de datos han reducido el tiempo invertido por las investigadoras en los procesos previos al análisis de la información. Sin embargo, los avances en estas líneas de investigación no solo se llevan a cabo a nivel nacional; el consorcio internacional de \textit{Wildlife Insights} \citep{ahumada2020wildlife} desarrolló una plataforma web que, además de la mayoría de las funciones de NAIRA, incluye la ventaja del almacenamiento de las imágenes en la nube. Adicionalmente, nuevos desarrollos se están evaluando en sistemas de IA local para transmitir casi en tiempo real la información de la identificación de la especie de estudio, empleando señales satelitales. 

\subsubsection*{Lecciones aprendidas}
El proceso que se ha venido desarrollando para la optimización de procesos alrededor del fototrampeo responde a una posición abierta al trabajo colaborativo y en red por lo que se sugiere como estrategia de investigación. Las futuras investigaciones pueden ampliar su enfoque de optimización en la diversificación de análisis empleando el mismo tipo de señales (visuales y acústicas) que se han venido recopilando. Asimismo, la eficiencia y precisión de la IA depende del entrenamiento que reciba. Esto hace que sea importante fortalecer los procesos que generen el mayor número de datos curados posible para luego ser incorporados en el entrenamiento de los algoritmos puesto que la recolección no es suficiente sin un proceso de curación con expertos. Bajo esta perspectiva la colaboración entre los expertos locales en los trópicos ayuda a mejorar los algoritmos de identificación de mamíferos, pero principalmente de aves, un grupo altamente diverso en los trópicos. 

\subsection*{2. Audición por computadora para el monitoreo y la conservación de la biodiversidad}

\subsubsection*{Contexto}
La emisión de señales acústicas es un comportamiento común en las especies terrestres, desempeñando roles diversos, como la defensa del territorio, el mantenimiento de la cohesión grupal, la atracción de parejas y la orientación \citep{fletcher2014}. Cada especie tiene sus propias señales distintivas, lo que proporciona información ecológica valiosa sobre la presencia, abundancia, estado y distribución de la comunidad animal \citep{sugai2019}. Las grabadoras de audio programables permiten registrar de manera autónoma y a lo largo de meses enteros el entorno sonoro, generando así una fuente invaluable de datos ecológicos. Sin embargo, el cuello de botella se presenta en el proceso de análisis ya que resulta ser lento y tedioso cuando se realiza de manera manual. La IA permite agilizar este flujo mediante el uso de modelos de escucha automatizada capaces de detectar y procesar estos datos de forma eficiente \citep{stowell2022}.

\subsubsection*{Solución propuesta}
Se ha llevado a cabo el desarrollo de rutinas computacionales que emplean aprendizaje supervisado para la identificación automática de vocalizaciones de especies de interés en Colombia (Figura \ref{fig:audio}). Tal es el caso del mono Leontocebus nigricollis en los bosques del Putumayo, así como múltiples especies de anfibios neotropicales en biomas como el Cerrado y el Bosque Atlántico \citep{canas2023}. Además, estos modelos han permitido detectar sonidos abióticos, como eventos de lluvia, que se desean identificar para su posterior exclusión en los análisis sobre estructura de la comunidad animal. En conjunto, estos modelos han posibilitado analizar grabaciones de campo de forma automatizada.
En la actualidad, el equipo del Instituto ha adaptado modelos pre-entrenados, como BirdNET \citep{kahl2021}, CLAP \citep{miao2023} y BatDetect2 \citep{mac2022}, para la identificación de marcas sonoras en las grabaciones. Además, se ha explorado la viabilidad de emplear modelos no supervisados y semi-supervisados para segmentar el paisaje sonoro en grupos de sonidos homogéneos, con el objetivo de simplificar el proceso de análisis y clasificación manual por expertos.
\subsubsection*{Lecciones aprendidas}
Una de las mayores necesidades al usar la IA en bioacústica es la de abordar los sesgos inherentes en los modelos de IA, por lo que se están desarrollando junto con expertas formas  rigurosas de evaluar su funcionamiento para evitar interpretaciones inadecuadas. Además, se ha evidenciado una falta de coordinación entre la comunidad de aprendizaje automático y la comunidad en ecología, enfatizando la importancia de una colaboración más estrecha. Por lo tanto, es evidente que no sólo las científicas de la computación desempeñan un papel importante en el desarrollo de estos modelos; tanto las expertas como las ecólogas, naturalistas y ciudadanas facilitan la recolección y revisión de datos diversos y anotados manualmente. Estos datos son insumos fundamentales para el desarrollo de modelos basados en aprendizaje automático. Además, estos cimientos sólidos son esenciales para el desarrollo continuo de algoritmos de IA efectivos en la conservación de la biodiversidad, y son de vital importancia en el Neotrópico, donde la biodiversidad está menos documentada que en ecosistemas templados.

\subsection*{3. Ciencia participativa en la IA}

\subsubsection*{Contexto}
Generalmente, se piensa que la ciencia es exclusiva para científicas y que solo se lleva a cabo en laboratorios y edificios de universidades e instituciones. No obstante, en la última década, la comunidad científica ha decidido abrir las puertas de la ciencia a la sociedad mediante ejercicios colaborativos con profesionales de otras áreas y personas con intereses diferentes. Es en este contexto donde surge la ciencia participativa, actuando como un puente de conocimiento entre las personas y asemejándose a la democracia, pues en ella cualquier persona puede tener voz y voto, ser responsable y participar de los procesos científicas. Diariamente, existe la preocupación generalizada por la calidad del ambiente, se monitorea el ritmo cardíaco y se disfruta fotografiando animales o paisajes. Estos intereses es lo que hace que cada persona pueda llegar a convertirse en potencial científico. Sin embargo, mucha de esta información se pierde cuando no se comparte. Es allí donde la tecnología desempeña un papel crucial en este puente de conocimiento, ya que actúa como catalizador de la información. Los teléfonos inteligentes y la IA pueden empoderar a las científicas ciudadanas y fortalecer una amplia red de conocimiento, ayudando a repensar cómo se genera el conocimiento, quién lo produce y a quién beneficia \citep{bonney2009}. 

\subsubsection*{Solución propuesta}
Las aplicaciones móviles como BirdNet, iNaturalist, Merlin, eBird, Waze, Mosquito Alert o Smoke Sense son ejemplos claros de cómo la tecnología transforma los datos generados por la sociedad en información pública curada. Todas estas aplicaciones operan de manera similar: ciudadanos voluntarios recopilan y registran información a través de ellas.
Si bien la ciencia ciudadana no tiene dentro de sus objetivos entrenar modelos de IA, la diversidad de sus datos y la cantidad que genera si tienen, o podrían llegar a tener, un impacto positivo en la mejora de sistemas de procesamiento especialmente en la naturalización del lenguaje o en la comprensión del contexto humano y científico en el que operan estos sistemas.
En la actualidad, algunas aplicaciones móviles y los proyectos de ciencia ciudadana están utilizando la IA, lo que ha a contribuido en el avance de la ciencia desde diferentes puntos de vista, entre ellos aumentar la velocidad, la calidad y la escala del procesamiento de los datos publicados, mejorar el alcance geográfico y temporal, y en especial, apoyar el entrenamiento de máquinas y diversificar las oportunidades de participación. 
En el procesamiento de datos, por ejemplo, las personas a través de aplicaciones como iNaturalist o eBird, etiquetan datos y los publican. Posteriormente, estos paquetes de datos y registros pueden ser usados para entrenar algoritmos y modelos de IA, y las personas pueden supervisar y validar los modelos para que puedan ser ajustados \citep{oliver2023a}. En cuanto a la mejora del alcance geográfico y temporal, los proyectos de ciencia participativa enfocados en monitoreos de tipo ambiental y climático dependen en gran medida de los datos que aportan las personas a través de sus dispositivos móviles. Con cada registro elaborado, las científicas ciudadanas ayudan a la IA usada para la predicción de eventos meteorológicos o que la identificación de especies sea cada vez más precisa \citep{wood2022}. 

\subsubsection*{Lecciones aprendidas}
La ciencia participativa representa una oportunidad enorme para mejorar el entrenamiento en aprendizaje de máquina y aumentar la capacidad de respuesta de la IA. La colección de datos usando ciencia ciudadana, es especialmente una oportunidad en lugares donde todavía no hay suficiente información sobre la biodiversidad, como lo es en el Neotrópico. También se ha identificado que se deben buscar formas de involucrar a las personas que están más cerca de la biodiversidad pero a la vez donde la infraestructura tecnológica es más limitada. Comunidades locales como indígenas o campesinos no sólo están más cerca de la biodiversidad sino que tienen un conjunto de valores y conocimientos que pueden mejorar las formas como podemos entender y actuar en la conservación \citep{pascual2023}.
Sin embargo, es clave cuestionarse a nivel ético y político como el manejo de datos de las científicas ciudadanas, que en su mayoría son voluntarias, pueden seguir siendo usados para el entrenamiento y cómo esto termina teniendo un impacto a nivel social, tecnológico y ambiental especialmente por que se desconoce cómo la IA puede garantizar la transparencia y la atribución de la información que está siendo usada. Asimismo es crucial pensar en las implicaciones de los sesgos intrínsecos de los datos (de genero, sexo, raciales, economicos, geograficos etc), que inherentemente serán replicados y amplificados por la IA.

\subsection*{4. La IA para el análisis de datos genéticos}

\subsubsection*{Contexto}

La identificación adecuada de especies y el entendimiento de sus interrelaciones dentro de un ecosistema son fundamentales para evaluar la biodiversidad y el estado de salud ambiental. El metabarcoding de ADN ambiental (eDNA) ha sur\-gido como una herramienta poderosa para detectar la presencia de múltiples especies a partir de muestras ambientales \citep{taberlet2018}. Sin embargo, los desafíos como la disponibilidad limitada de bases de datos de referencia y el procesamiento de grandes volúmenes de datos complejos pueden obstaculizar una identificación taxonómica confiable \citep{deiner2017}. Por otro lado, a medida que las técnicas de secuenciación avanzan, la cantidad de datos generados supera la capacidad de análisis manual, lo que resalta la necesidad de soluciones automatizadas y escalables \citep{fluck2022}.

\subsubsection*{Solución propuesta}
El uso de técnicas de IA ha demostrado un gran potencial para abordar desafíos en el análisis de datos de metabarcoding. Por ejemplo, la aplicación de enfoques de aprendizaje de máquina, como las redes neuronales profundas y el aprendizaje automático supervisado, ha mejorado significativamente la precisión en la identificación taxonómica de organismos a partir de códigos de barras de ADN y datos de ADN ambiental o eDNA \citep{kimura2022,lamperti2023}. En particular, un marco robusto de aprendizaje automático ha sido propuesto para clasificar secuencias de eDNA en ambientes complejos, ayudando a desentrañar patrones ocultos en la biodiversidad \citep{yang2024}. Además, el uso de métodos como el "eco-tag" ha permitido manejar la variabilidad genética intraespecífica mediante el uso de IA, mejorando la identificación de especies y la resolución de unidades taxonómicas moleculares operativas \citep{boyer2016,allard2023}.
Los métodos basados en aprendizaje profundo, como las redes neuronales convolucionales (CNNs), también han demostrado ser eficaces en la asignación taxonómica de secuencias de eDNA en ecosistemas complejos como los marinos, proporcionando una mayor precisión y velocidad en comparación con los métodos tradicionales \citep{fluck2022}. Además, las nuevas metodologías de aprendizaje profundo, como las basadas en autoencoders y aprendizaje métrico, han permitido visualizar y analizar patrones de diversidad en datos de eDNA con mayor detalle, extrayendo características ecológicas relevantes que no eran detectables con técnicas convencionales \citep{lamperti2023}.
La aplicación de técnicas de IA como redes neuronales profundas y procesamiento de lenguaje natural (NLP) en eDNA no solo facilita la identificación precisa de especies, sino también la revelación de patrones y relaciones ecológicas complejas. Por ejemplo, los modelos de IA entrenados con datos de eDNA y características ambientales pueden predecir la presencia y distribución de especies bioindicadoras, proporcionando información crítica sobre el estado de salud de los ecosistemas \citep{muha2017, rahim2023}. Estas capacidades predictivas son fundamentales para comprender sus dinámicas y desarrollar estrategias efectivas de conservación \citep{bessey2021, hassan2022}

\subsubsection*{Lecciones aprendidas}
Si bien aún existen desafíos, como la disponibilidad y calidad de los datos de entrenamiento y la interpretación de los modelos de IA, el potencial de estas técnicas para mejorar la identificación de especies y desentrañar las relaciones ecológicas es prometedor. Por ejemplo, la integración de herramientas de aprendizaje profundo ha permitido la creación de modelos híbridos que combinan información genética y ambiental para una clasificación más precisa de comunidades biológicas \citep{lamperti2023}. La aplicación de redes neuronales convolucionales ha acelerado significativamente el procesamiento de secuencias de eDNA, facilitando la automatización del monitoreo en ecosistemas complejos \citep{fluck2022}. La integración efectiva del metabarcoding y la IA podría revolucionar nuestra comprensión de la biodiversidad y facilitar el monitoreo y la protección de los ecosistemas a largo plazo, permitiendo un enfoque más robusto y dinámico para enfrentar los desafíos ambientales actuales.

\subsection*{5. Usos de la IA para el fortalecimiento de sistemas de información sobre biodiversidad}

\subsubsection*{Contexto}
Los sistemas de información sobre biodiversidad tienen como propósito brindar acceso abierto a información sobre diversidad biológica, facilitando la publicación en línea de datos e información y promoviendo su uso por parte de una amplia variedad de audiencias. Adicionalmente, apoya de forma oportuna y eficiente la gestión integral de la biodiversidad al proporcionar datos que soportan la toma de decisiones informada a diferentes escalas, así como procesos de investigación y educación que alcanzan públicos diversos. Estos esfuerzos a nivel nacional o temático se articulan en la plataforma global de información sobre biodiversidad (GBIF por sus siglás en inglés), disponible en: \url{www.gbif.org}, dándoles mayor visibilidad e impacto. Para potenciar algunas de las líneas de trabajo de estos sistemas, es posible utilizar la IA en la gestión y fortalecimiento de capacidades de las organizaciones que participan en ellos, facilitando la digitalización y publicación de datos, y mejorando la calidad de los que ya han sido publicados (Figura \ref{fig:Ddata}).

\subsubsection*{Solución propuesta}

\begin{itemize}
    \item Gestión de organizaciones que participan en los sistemas: la IA ha apoyado análisis de bases de datos identificando patrones, segmentando posibles socios, agilizando la comunicación y proporcionando respuestas personalizadas aprendiendo con cada interacción \citep{burns2023}. Sin embargo, existen riesgos como la dificultad de mantener socios vinculados a largo plazo \citep{miller2022} o la perpetuación de sesgos sociales que pueden generar discriminación \citep{dastin2022}. En este sentido, ha sido clave mantener principios claros de transparencia, equidad y privacidad, dándoles la oportunidad a todos los involucrados con la IA de mantener la interacción con ella o no \citep{jobin2019, ryan2021}.
    \item Fortalecimiento de colecciones biológicas: la IA puede facilitar y mejorar el flujo de trabajo entre colecciones biológicas y los sistemas de información al permitir la identificación automática de ejemplares comunes a partir de imágenes o sonidos \citep{schermer2018, mendoza2023past} o apoyando la digitalización de etiquetas de especímenes a partir de la toma de fotografías desde múltiples ángulos y el uso de redes neuronales convolucionales \citep{salili2023}. Estas tareas ocupan tiempo de los taxónomos, lo que retrasa la publicación de datos valiosos en sistemas de información accesibles para su uso en investigación, educación y toma de decisiones ambientales \citep{stenhouse2023}.
    \item Validación y calidad de datos: después de la digitalización y publicación de los datos es necesario seguir evaluando su calidad y consistencia. En los últimos años, diferentes investigaciones enfocadas en el análisis de datos han implementado técnicas de aprendizaje supervisado y no supervisado con el fin de identificar la duplicidad de registros, inconsistencias y facilitar el descubrimiento de nuevas especies. Dentro de los ejemplos que se han aplicado a datos biológicos publicados en GBIF está el algoritmo no supervisado DBCSCAN, el cual emplea información de colectores, fechas y número de registro con el fin de determinar las expediciones donde más especies se han descubierto \citep{nicolson2017} o la detección de valores geográficos atípicos \citep{waller2020}. Adicionalmente, se pueden implementar diferentes estudios con los datos finales publicados; por ejemplo, el algoritmo de Random forest se ha usado como una herramienta que permite identificar posibles especies hospederas y puntos geográficos claves para un organismo patógeno \citep{robles2022}.
\end{itemize}

\subsubsection*{Lecciones aprendidas}
Se espera que los sistemas de información sobre biodiversidad que incorporen la IA en su gestión de socios, puedan relacionarla directamente con ellos de manera transparente, entrenando modelos para dar respuestas cada vez más personalizadas y que conecten con sus motivaciones. Siempre con el enfoque de enriquecer las interacciones y no sólo reducir costos o aumentar la eficiencia \citep{burns2023}. En cuanto a la gestión, digitalización y calidad de datos, se espera realizar la implementación de diferentes rutinas que permitan automatizar los procesos e identificar rápidamente inconsistencias que puedan ser solucionadas en tiempo real.

\subsection*{6. IA para el modelamiento de la distribución geográfica de la biodiversidad}

\subsubsection*{Contexto}

El conocimiento sobre las distribuciones geográficas de las especies en escenarios pasados,  presentes y futuros es esencial para definir estrategias de planificación del territorio y priorizar áreas que tienen mayor probabilidad de apoyar la adaptación al cambio climático. La Red de Observación de la Biodiversidad de Colombia (Colombia BON) ha documentado avances en la integración de modelos de distribución de especies con información satelital que sea relevante para los tomadores de decisiones. Sin embargo, la mayoría de especies no tiene mapas de distribución, y para aquellas que sí tienen, suelen ser imprecisos, desactualizados y de difícil acceso para muchos usuarios finales. Es por ello que a través de la plataforma BioModelos \citep{velasquez2019} se han desarrollado avances en el modelado para representar hábitats de manera estática, lo que captura parcialmente la dinámica completa de las especies y sus entornos cambiantes. Esto puede limitar su impacto en la toma de decisiones informada y adaptativa frente a desafíos como el cambio climático.
\subsubsection*{Solución propuesta}
En los últimos años se han desarrollado herramientas de apoyo para integrar, procesar y analizar información de distribución de especies con observaciones de la Tierra, como datos climáticos y de hábitat. Estas herramientas integran algoritmos de aprendizaje automático (Support Vector Machine, Boosted Regression Trees, Random Forest) \citep{hastie2009elements} y modelos de máxima entropía (MaxEnt) \citep{phillips2008} para mejorar la estimación de la distribución de especies. Algunas de éstas han sido modeladas con parámetros que reflejan todas las combinaciones de variables ambientales y parámetros de optimización \citep{muscarella2014}. A partir de estos modelos se han obtenido, al menos, 8700 mapas de modelos de distribución de especies, los cuales fueron integrados en un servicio web que usa información climática y de biodiversidad, para apoyar las decisiones de las agencias colombianas hacia la identificación de estrategias de conservación complementarias en el país (ver página web: \url{https://biomodelos.humboldt.org.co/}). Paralelamente, a través de Google Earth Engine, usando algoritmos de aprendizaje de máquina (supporting vector machine, MaxEnt, Random Forest y Boosted Regression Trees), se desarrollaron modelos dinámicos, específicamente para aves, que capturan los cambios en la disponibilidad de hábitat en las últimas dos décadas (2000-2021), dando pistas sobre las tendencias de cambio de hábitat. Sin embargo, ha sido necesario complementar la evaluación de los mapas de hábitat considerando que algunos de ellos muestran errores de predicción en los límites de las áreas de distribución. Adicionalmente se implementó un proceso participativo con expertos, filtrando así el mejor modelo para cada especie. Los mapas filtrados por criterio experto y por valores de fiabilidad >= 70\% están disponibles en línea ( \url{https://biomodelos-iavh.users.earthengine.app/view/biomodelos}).
\subsubsection*{Lecciones aprendidas}
El uso de aprendizaje automático y geoprocesamiento en la nube acelera significativamente los tiempos de modelado, permitiendo ejecutar múltiples modelos y probar diferentes metodologías simultáneamente. Sin embargo, no es recomendable usar un único conjunto de parámetros, variables y escalas para todas las especies, ya que cada una tiene características particulares. Por ello, es crucial combinar los modelos con evaluación experta. Además, los modelos deben explicar las bases de sus predicciones para ser más confiables. Estos mapas, que resultan de un modelo de aprendizaje de máquina y evaluación por expertos, contribuyen a generar alertas de pérdida de hábitat, mejorando los esfuerzos de conservación a diversas escalas.

\subsection*{7. La IA como herramienta para impulsar la bioeconomía sostenible en Colombia}

\subsubsection*{Contexto}
 
A lo largo de las últimas décadas, se ha discutido ampliamente sobre la riqueza que posee Colombia en términos de sus recursos naturales. Sin embargo, esta diversidad aún no representa un factor de competitividad para el país para cada una de sus regiones \citep{maldonado2020}, debido a la falta de identificación detallada de los componentes específicos de la biodiversidad y la especificidad geográfica de las especies \citep{gori2022understanding}. Por ello, se ha venido promoviendo el uso sostenible de la biodiversidad y sus servicios ecosistémicos (SSEE) como una vía para avanzar hacia un modelo de desarrollo más armonioso con el entorno que permita impulsar de forma paralela la conservación de la biodiversidad y la gestión integral de los recursos que lo componen. En este contexto, la IA se identifica como una herramienta para fortalecer dicha gestión de la biodiversidad, contribuyendo a la conservación y a la promoción de un bienestar humano que esté en consonancia con la salud del ecosistema.

\subsubsection*{Solución propuesta}
Algunos autores como \citep{sodergard2021} discuten cómo la IA juega un papel preponderante en materia de monitoreo (p.ej.: trazabilidad en procesos), optimización (p.ej.: productividad, tiempos y costos) y predicción (p.ej.: características funcionales) en cualquier etapa o proceso que asocie estas actividades. El rango de actuación se amplía dadas las necesidades particulares de cada línea estratégica que pueda ser definida en el marco del uso eficiente de los recursos naturales (Tabla \ref{tab:tabla_esp}). Además de estos referentes, existen otras aplicaciones a nivel mundial que podrían ser específicamente propuestas para abordar las necesidades relacionadas con cada una de ellas.

\subsubsection*{Lecciones aprendidas}
El papel de la IA en la promoción de alternativas que propendan por el uso y aprovechamiento de la biodiversidad habilita la incorporación, potencialización y fortalecimiento de sistemas productivos socio-ecológicos (i.e. convergencia entre conservación y productividad) en un país con una alta heterogeneidad de contextos y desafíos por región. Lo anterior se vuelve relevante en el proceso de transición gradual y transicional hacia la bioeconomía que se encuentra atravesando Colombia, ya que los resultados, producto de la aplicación de la herramienta, habilitan un desarrollo más equitativo y competitivo al brindar facilidades en contextos donde la integración de diversas fuentes de conocimiento es necesaria. No obstante, uno de los retos de su aplicación implica el reconocimiento de la adopción de estas herramientas a las necesidades específicas, dada la amplia gama de posibilidades que ofrece para agregar valor a los procesos asociados al uso eficiente de los recursos naturales. Además, se destaca la oportunidad que proporciona para realizar acciones en el corto y mediano plazo, que no solo sean solo conceptuales o indirectas, sino que también permita ver a la biodiversidad  como un factor habilitante para el país.

\subsection*{8. Optimización del aprovechamiento soste\-nible de la biodiversidad colombiana: el papel de la IA y LangChain en la Bioprospección In Silico}

\subsubsection*{Contexto}
La bioprospección es la búsqueda sistemática y exploración de material biológico y fuentes naturales de moléculas e información bioquímica y genética, con el potencial de generar productos valiosos para diferentes sectores \citep{harvey2011}. Este proceso implica no solo la investigación de los recursos naturales con el fin de descubrir nuevos compuestos, sino también cultivar y aislar las especies que puedan producir dichos compuestos de interés, seguido de su validación experimental \citep{kamble2022}. La bioprospección ha sido una de las grandes apuestas en Colombia \citep{dnp2018} ya que desde el Convenio de Diversidad Biológica (CDB), se ha reconocido como un mecanismo para alcanzar los objetivos propuestos de conservación y uso de la biodiversidad \citep{naciones2007}. Esta apuesta se fundamenta en que, a partir del conocimiento de la biodiversidad, es posible generar productos comercialmente valiosos y sostenibles que aporten en la transición hacia modelos más sostenibles, como la bioeconomía, y la consolidación de negocios verdes y economías locales. Si bien el proceso proporciona un mecanismo para hacer aprovechamiento de la biodiversidad, el enfoque convencional suele considerarse ineficiente en términos de costo y tiempo, ya que depende, entre otros, de la pureza y calidad de las muestras \citep{kamble2022}. Es allí donde surgen nuevos enfoques como la bioprospección in silico que hacen uso de herramientas computacionales para acelerar el proceso de investigación relacionado con la búsqueda de componentes bioactivos presentes en el bioma. La bioprospección in silico implica la identificación de nuevos candidatos para la generación de productos utilizando bases de datos y herramientas informáticas para el cribado virtual de la información biológica. Este método se considera simple, eficaz y menos costoso, y suele estar mediado por dos pasos: la exploración de bases de datos; y después, la caracterización in silico, que implica la selección, análisis y preselección de los nuevos candidatos utilizando herramientas bioinformáticas.
\subsubsection*{Solución propuesta}
Aplicar la bioprospección in silico  \citep{santana2021}, como una aproximación mediante IA que reduzca la inversión monetaria y los esfuerzos realizados tradicionalmente, sería un punto de partida para un mayor entendimiento de los datos biológicos y su proyección dentro de un contexto de sostenibilidad. Específicamente, la IA permite el uso de datos pre-existentes generados convencionalmente y conlleva a la exploración de esa información para la predicción de familias de metabolitos con funciones biológicas relevantes. Particularmente, la IA requiere de herramientas complementarias en el análisis de datos para abordar interrogantes antes de llevar a cabo un análisis en campo. En este contexto, el LangChain emerge como una herramienta crucial al proporcionar modelos para agilizar la identificación de especies con potencial prometedor, con el fin último de orientar y acelerar el proceso de bioprospección tradicional al focalizar los esfuerzos y permitir que se tomen decisiones informadas. El objetivo es desarrollar una metodología base empleando estas aproximaciones, para hacer eficiente la ruta de selección de especies nativas colombianas con potencial en bioeconomía, considerando que la integración de herramientas computacionales y datos biológicos puede mejorar la recopilación de información. Además, habilita la identificación de compuestos bioactivos valiosos con aplicación en diversos sectores, para posteriormente ser validados en procesos de bioprospección tradicional. 
\subsubsection*{Lecciones aprendidas}
Se reconoce la oportunidad de aprovechar los recursos naturales de Colombia de manera sostenible partiendo del uso de la IA como perspectiva innovadora y accesible para potenciar la bioeconomía del país. Esta visión implica integrar la IA como apoyo al proceso de bioprospección tradicional, reduciendo la cantidad de inversión de recurso humano y capital en estos análisis y como un prometedor enfoque para identificar especies nativas con potencial basándose en su composición bioquímica. Esta iniciativa no solo destaca el potencial de la herramienta sino que también agiliza la selección de las especies como mecanismo para impulsar el bienestar humano y la conservación de manera simultánea. 
Así, para fortalecer el trabajo en campo, se propone el uso de herramientas in silico, que permitan imaginar y crear nuevas posibilidades, dinámicas y relaciones, respetando los límites ecológicos establecidos en el contexto de la resiliencia planetaria. Además, se busca adaptar medidas efectivas para detener la pérdida de diversidad biológica, mientras se fomentan alternativas económicas que contribuyan e incorporen la biodiversidad como factor clave para mejorar el bienestar humano.

\switchcolumn

\section*{Experiences and applications}

This section presents case studies and applications of artificial intelligence in the field of conservation and sustainable use of biodiversity. This is done through practical examples from various perspectives, including biodiversity monitoring, modeling of biological records, bioeconomics, genetic data analysis, among others. Each case offers a specific context, demonstrating how AI provides innovative solutions and what lessons learned can be highlighted from its implementation. Thus, based on these studies, a literature review and team discussions, the main challenges and future directions to consider have been identified.  
\\
\subsection*{1. Camera trapping and AI in nature conservation}

\subsubsection*{Context}

Cryptic mammal and bird species with small population sizes have been challenging to study until camera traps allowed for reduced human presence in obtaining photographic records (of cryptic species) associated with spatial and temporal metadata, among others. As technology has advanced, the data collected has exponentially increased; current devices can capture images for at least a month without human presence \citep{oliver2023a}. This has lowered costs, allowing a greater number of devices to be used in each study. Given this data landscape, efficient management is essential to optimize the processes required for analysis and draw conclusions that support decision-making for nature conservation.

\subsubsection*{Proposed solution}
Interdisciplinary and collaborative work is integral to the research strategy that has successfully advanced process optimization. The Humboldt Institute has created collaborations to create a desktop application called NAIRA, now in its third version in partnership with the University of Antioquia. NAIRA III is designed to facilitate the post-processing of images obtained from camera traps, allowing users to classify images by identifying the animals present in each photo and separating those without wildlife \citep{giraldo2019, pulido2018}. This version automatically identifies photos containing birds and mammals, including 20 taxonomic genera within the latter group. It also automatically extracts information from the photos, such as moon phase, temperature, GPS coordinates, date, and time of capture, organizing it into tabular data following the Darwin Core standard. In addition, it automatically calculates a table of independent records along with presence-absence matrices and counts by dates, moon phases, and temperature. Another feature of the program allows users to manually add important information to analyze each camera trap session and to modify the automatically obtained classifications.
AI and data organization systems have reduced the time researchers spend on processes prior to data analysis. However, advances in these research areas are not limited to national efforts; the international \textit{Wildlife Insights} consortium \citep{ahumada2020wildlife} has developed a Web platform that, in addition to most NAIRA functions, offers cloud storage for images. Furthermore, new developments are being assessed in local AI systems to transmit species identification information almost in real time using satellite signals.

\subsubsection*{Lessons learned}
The process that has been developed to optimize camera trapping workflows reflects an open approach to collaborative and networked work, which is proposed as a research strategy. Future research can expand its optimization focus by diversifying analyses using the same types of signals (visual and acoustic) that have been collected so far. Additionally, the efficiency and accuracy of AI depend on the training it receives. This makes it crucial to strengthen processes that generate as much curated data as possible, which can then be incorporated into algorithm training, as data collection alone is insufficient without expert-led curation. From this perspective, collaboration with local experts in the tropics helps improve identification algorithms for mammals, and especially for birds, a highly diverse group in tropical regions.

\begin{figure}[h]
    \centering
    \includegraphics[width=0.5\textwidth]{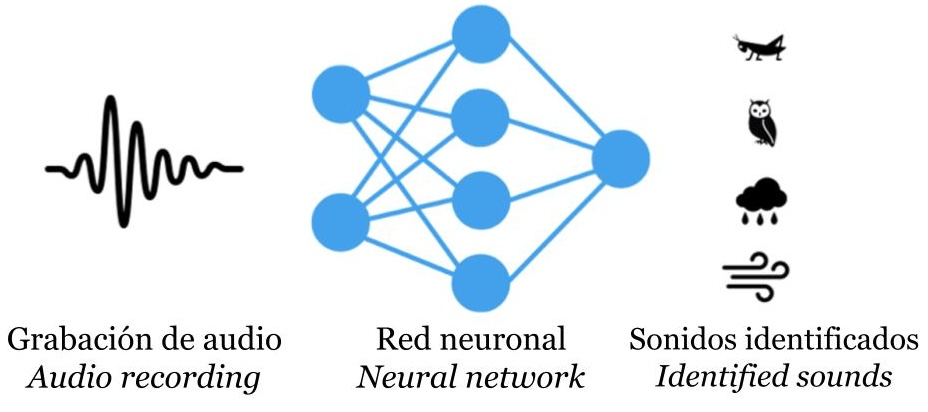}
    \caption{Imágen ilustrativa que representa  cómo el aprendizaje de máquina puede ayudar a descomponer y anotar los principales elementos del paisaje sonoro. \textit{Illustrative image representing how machine learning can help break down and annotate the main elements of the soundscape.}}
    \label{fig:audio}
\end{figure}

\subsection*{2. Machine Listening for monitoring and conserving biodiversity}

\subsubsection*{Context}
The emission of acoustic signals is a common behavior in terrestrial species, which plays different roles, such as territorial defense, group cohesion maintenance, attraction of partners, and orientation \citep{fletcher2014}. Each species has its own distinctive signals, providing valuable ecological information about the presence, abundance, status, and distribution of the animal community \citep{sugai2019}. Programmable audio recorders can autonomously record the sound environment over several months, generating an invaluable source of ecological data. However, a bottleneck occurs during the analysis process, which can be slow and tedious when conducted manually. AI can streamline this workflow through automated listening models capable of efficiently detecting and processing these data \citep{stowell2022}.   
\\
\subsubsection*{Proposed solution}
Development has taken place in Colombia of computational routines that employ supervised learning for the automatic identification of vocalizations from species of interest (Figure  \ref{fig:audio}). This includes the Leontocebus nigricollis monkey in the Putumayo forests, as well as multiple species of Neotropical amphibians in biomes such as the Cerrado and Atlantic Forest \citep{canas2023}. Furthermore, these models have allowed for the detection of abiotic sounds, such as rainfall events, which are identified for later exclusion from analyses on the structure of the animal community. Together, these models enable automated analysis of field recordings.
Currently, the Institute Humboldt bioacoustics team has adapted pretrained models, such as BirdNET \citep{kahl2021}, CLAP \citep{miao2023} and, BatDetect2 \citep{mac2022} to identify sound markers in recordings. Additionally, the feasibility of using unsupervised and semi-supervised models to segment the soundscape into groups of homogeneous sounds has been explored, aiming to simplify the manual analysis and classification process by experts.

\subsubsection*{Lessons learned}
One of the main needs when using AI in bioacoustics is to address the inherent biases in AI models. Therefore, rigorous evaluation methods are being developed in collaboration with experts to prevent misinterpretations. Furthermore, a lack of coordination has been observed between the machine learning community and the ecology community, highlighting the importance of closer collaboration. It is clear that computer scientists are not the only ones playing a crucial role in developing these models; experts in ecology, naturalists, and citizen scientists facilitate the collection and review of various manually annotated data. These data are a fundamental input for the development of machine learning models. Moreover, these solid foundations are essential for the continued development of effective AI algorithms for biodiversity conservation, particularly in the Neotropics, where biodiversity is less documented than in temperate ecosystems. 
\\

\subsection*{3. Participatory science in AI}

\subsubsection*{Context}

Science is generally believed to be exclusively for scientists and that it takes place only in laboratories and university buildings. However, in the last decade, the scientific community has decided to open the doors of science to society through collaborative efforts with professionals in other fields and individuals with diverse interests. This is where participatory science emerges, acting as a bridge of knowledge between people and resembling democracy, as anyone can have a voice and vote, responsible, and participate in scientific processes. There is a widespread daily concern for environmental quality, monitoring heart rates, and enjoying photographing animals or landscapes. These interests mean that anyone can potentially become a scientist. However, much of this information is lost when it is not shared. This is where technology plays a crucial role in this bridge of knowledge, acting as a catalyst for information. Smartphones and AI can empower citizen scientists and strengthen a wide knowledge network, helping to rethink how knowledge is generated, who produces it, and who benefits from it \citep{bonney2009}.
\\
\\

\subsubsection*{Proposed solution}
Mobile applications like BirdNet, iNaturalist, Merlin, eBird, Waze, Mosquito Alert, and Smoke Sense are clear examples of how technology transforms data generated by society into curated public information. All these applications operate similarly: citizen volunteers collect and register information through them.
Although citizen science does not aim to train AI models, the diversity of its data and the quantity generated do have, or could have, a positive impact on improving processing systems, especially in natural language understanding and in comprehending the human and scientific context in which these systems operate.
Currently, some mobile applications and citizen science projects are using AI, contributing to scientific advancement from various perspectives, including increasing the speed, quality, and scale of published data processing, improving geographic and temporal reach, and, importantly, supporting machine training and diversifying participation opportunities.
In data processing, for example, individuals use applications like iNaturalist or eBird to tag data and publish it. Subsequently, these data packets and records can be used to train algorithms and AI models, and individuals can supervise and validate the models so that they can be adjusted \citep{oliver2023a}. Regarding improving geographic and temporal reach, participatory science projects focused on environmental and climate monitoring are largely based on data provided by individuals through their mobile devices. With each record created, citizen scientists help improve the AI used to predict weather events or for more accurate species identification \citep{wood2022}.\\
\\
\\

\subsubsection*{Lessons learned}
Participatory science represents a tremendous opportunity to improve machine learning training and increase AI response. Collecting data through citizen science is especially valuable in places where there is still insufficient information about biodiversity, such as the Neotropics. It has also been identified that efforts should be made to involve people closest to biodiversity, where technological infrastructure is often limited. Local communities, such as Indigenous peoples or farmers, are not only closer to biodiversity but also possess a set of values and knowledge that can improve how we understand and act in conservation \citep{pascual2023}.
However, it is crucial to ethically and politically question how data from citizen scientists, who are mostly volunteers, can continue to be used for training and how this impacts social, technological, and environmental levels, especially since it is unknown how AI can ensure the transparency and attribution of the information being used. Additionally, it is essential to consider the implications of intrinsic biases in the data (gender, sexual, racial, economic, geographical, etc.) that will inherently be replicated and amplified by AI.

\subsection*{4. AI for genetic data analysis}

\subsubsection*{Context}

Accurate species identification and understanding their interrelations within an ecosystem are fundamental to assessing biodiversity and environmental health. Environmental DNA metabarcoding (eDNA) has emerged as a powerful tool to detect the presence of multiple species in environmental samples \citep{taberlet2018}. However, challenges such as the limited availability of reference databases and the processing of large volumes of complex data can hinder reliable taxonomic identification \citep{deiner2017}. As sequencing techniques advance, the amount of generated data exceeds the capacity for manual analysis, highlighting the need for automated and scalable solutions \citep{fluck2022}.
\\
\subsubsection*{Proposed solution}
The use of AI techniques has shown great potential in addressing challenges in metabarcoding data analysis. For example, the application of machine learning approaches, such as deep learning and supervised machine learning, has significantly improved the accuracy of taxonomic identification of organisms from DNA barcodes and eDNA data \citep{kimura2022,lamperti2023}. In particular, a robust machine learning framework has been proposed to classify eDNA sequences in complex environments, helping to uncover hidden patterns in biodiversity \citep{yang2024}. Additionally, methods like "eco-tag" have managed intra-specific genetic variability using AI, improving species identification and resolving operational taxonomic units \citep{boyer2016,allard2023}.
Deep learning methods, such as convolutional neural networks (CNNs), have also proven effective in the taxonomic assignment of eDNA sequences in complex ecosystems such as marine environments, providing greater accuracy and speed compared to traditional methods \citep{fluck2022}. Furthermore, new deep learning methodologies, such as those based on autoencoders and metric learning, have allowed more detailed visualization and analysis of diversity patterns in eDNA data, extracting relevant ecological features that were undetectable with conventional techniques \citep{lamperti2023}.
The application of AI techniques, such as deep neural networks and natural language processing (NLP) in eDNA analysis, not only facilitates precise species identification but also reveals complex ecological patterns and relationships. For example, AI models trained with eDNA data and environmental features can predict the presence and distribution of bioindicator species, providing critical information about ecosystem health \citep{muha2017, rahim2023}. These predictive capabilities are essential for understanding dynamics and developing effective conservation strategies \citep{bessey2021, hassan2022}.

\subsubsection*{Lessons learned}
While challenges remain, such as the availability and quality of training data and the interpretability of AI models, the potential of these techniques to improve species identification and unravel ecological relationships is promising. For example, integrating deep learning tools has led to the creation of hybrid models that combine genetic and environmental information for more accurate classification of biological communities \citep{lamperti2023}. The application of convolutional neural networks has significantly accelerated the processing of eDNA sequences, facilitating the automation of monitoring in complex ecosystems \citep{fluck2022}. The effective integration of metabarcoding and AI could revolutionize our understanding of biodiversity and enhance long-term monitoring and protection of ecosystems, allowing for a more robust and dynamic approach to addressing current environmental challenges.

\subsection*{5. Uses of AI to strengthen biodiversity information systems}

\subsubsection*{Context}

Biodiversity information systems aim to provide open access to information on biological diversity, facilitate online publication of data and information, and promote its use by a wide variety of audiences. In addition, they efficiently support the comprehensive management of biodiversity by providing data that underpin informed decision-making on different scales, as well as research and educational processes that reach diverse public. These national or thematic efforts are integrated into the global biodiversity information platform (GBIF), available at \url{www.gbif.org}, enhancing their visibility and impact. To enhance some of the lines of work of these systems, AI can be utilized to manage and strengthen the capacities of the organizations involved, facilitating data digitization and publication, and improving the quality of published information (Figure \ref{fig:data_eng}).

\subsubsection*{Proposed solution}

\begin{itemize}
    \item Management of organizations participating in the systems: AI has supported database analyses by identifying patterns, segmenting potential partners, streamlining communication, and providing personalized responses that learn from each interaction \citep{burns2023}. However, there are risks, such as the challenge of maintaining long-term partnerships \citep{miller2022} and the perpetuation of social biases that may lead to discrimination \citep{dastin2022}. In this regard, it has been crucial to uphold clear principles of transparency, equity, and privacy, allowing all stakeholders involved with AI the option to engage with it or not \citep{jobin2019, ryan2021}.

    \item Strengthening biological collections: AI can facilitate and improve workflow between biological collections and information systems by enabling the automatic identification of common specimens from images or sounds \citep{schermer2018,mendoza2023past} or supporting the digitization of specimen labels through photographs taken from multiple angles using convolutional neural networks \citep{salili2023}. These tasks consume taxonomists' time, delaying the publication of valuable data in information systems accessible for research, education, and environmental decision-making \citep{stenhouse2023}.
    \item Data validation and quality: After digitizing and publishing data, it is necessary to continually assess their quality and consistency. In recent years, various researchers focused on data analysis have implemented supervised and unsupervised learning techniques to identify duplicate records, inconsistencies, and facilitate the discovery of new species. Examples applied to biological data published in GBIF include the unsupervised algorithm DBSCAN, which uses collector information, dates, and record numbers to determine expeditions with the highest species discoveries \citep{nicolson2017}, or detecting anomalous geographic values \citep{waller2020}. Additionally, various studies can be conducted with the final published data; for example, the Random Forest algorithm has been used as a tool to identify potential host species and key geographic points for a pathogenic organism \citep{robles2022}.
\end{itemize}

\subsubsection*{Lessons learned}
It is expected that biodiversity information systems incorporating AI in their partner management will be able to relate directly and transparently to them, training models to provide increasingly personalized responses that connect with their motivations. The focus should always be on enriching interactions rather than merely reducing costs or increasing efficiency \citep{burns2023}. Regarding management, digitization, and data quality, the aim is to implement various routines that allow for the automation of processes and the rapid identification of inconsistencies that can be resolved in real time.

\subsection*{6. AI for Modeling the Geographic Distribution of Biodiversity}

\subsubsection*{Context}

Knowledge about the geographic distributions of species in past, present, and future scenarios is essential for defining land planning strategies and prioritizing areas that are most likely to support adaptation to climate change. The Colombian Biodiversity Observation Network (Colombia BON) has documented progress in integrating species distribution models with satellite information that is relevant for decision-makers. However, most species do not have distribution maps, and for those that do, the maps tend to be inaccurate, outdated, and difficult for many end users to access. Therefore, through the BioModelos platform \citep{velasquez2019}, advancements have been made in the modeling process to represent habitats statically, which only partially captures the full dynamics of species and their changing environments. This can limit its impact on informed and adaptive decision-making in the face of challenges such as climate change.

\subsubsection*{Proposed solution}
In recent years, support tools have been developed to integrate, process, and analyze species distribution information with Earth observations, such as climate and habitat data. These tools integrate machine learning algorithms (Support Vector Machine, Boosted Regression Trees, Random Forest) \citep{hastie2009elements} and maximum entropy models (MaxEnt) \citep{phillips2008} to improve species distribution estimation. Some of these have been modeled with parameters reflecting all combinations of environmental variables and optimization parameters \citep{muscarella2014}. From these models, at least 8,700 species distribution model maps have been obtained, which were integrated into a web service that uses climate and biodiversity information to support decisions by Colombian agencies in identifying complementary conservation strategies in the country (see website: \url{https://biomodelos.humboldt.org.co/}). Simultaneously, through Google Earth Engine, using machine learning algorithms (support vector machine, MaxEnt, Random Forest, and Boosted Regression Trees), dynamic models were developed specifically for birds, capturing changes in habitat availability over the past two decades (2000-2021), providing insights into habitat change trends. However, it has been necessary to complement the evaluation of habitat maps considering that some of them show prediction errors at the distribution area boundaries. Additionally, a participatory process with experts was implemented to filter the best model for each species. The maps filtered by expert criteria and by reliability values bigger than 70 \% are available online (\url{https://biomodelos-iavh.users.earthengine.app/view/biomodelos}).

\subsubsection*{Lessons learned}
The use of machine learning and cloud geoprocessing significantly accelerates modeling times, allowing the execution of multiple models and testing of different methodologies simultaneously. However, it is not advisable to use a single set of parameters, variables, and scales for all species, as each has its own particular characteristics. Therefore, it is crucial to combine models with expert evaluation. Additionally, the models should explain the basis for their predictions to be more reliable. These maps, resulting from machine learning modeling and expert assessment, help generate habitat loss alerts, improving conservation efforts at various scales.

\subsection*{7. AI as a Tool to Promote Sustainable Bioeconomy in Colombia}
\subsubsection*{Context}

Throughout the last decade, there has been extensive discussion about Colombia's wealth in terms of its natural resources. However, this diversity has yet to translate into a competitive advantage for the country in each of its regions \citep{maldonado2020}, due to the lack of detailed identification of specific components of biodiversity and the geographical specificity of species \citep{gori2022understanding}. Therefore, sustainable use of biodiversity and its ecosystem services (ESS) has been promoted as a pathway toward a development model that is more harmonious with the environment, allowing simultaneous advancement of biodiversity conservation and integrated management of its resources. In this context, artificial intelligence is identified as a tool to strengthen biodiversity management, contributing to conservation, and promoting human well-being that aligns with ecosystem health.
\subsubsection*{Proposed solution}
Authors like \citep{sodergard2021} discuss how AI plays a significant role in monitoring (e.g., traceability in processes), optimization (e.g., productivity, time, and costs), and prediction (e.g., functional characteristics) at any stage or process associated with these activities. The scope of action expands given the particular needs of each strategic line that can be defined within the framework of efficient use of natural resources (Table \ref{tab:tabla_eng}). In addition to these references, there are other applications globally that could be specifically proposed to address the needs related to each of them.

\subsubsection*{Lessons learned}
The role of AI in promoting alternatives for the sustainable use and exploitation of biodiversity enables the incorporation, enhancement, and strengthening of socio-ecological production systems (i.e., convergence between conservation and productivity) in a country with high heterogeneity of contexts and challenges by region. This becomes relevant in the gradual and transitional process towards a bioeconomy that Colombia is undergoing, as the results of applying the tool enable a more equitable and competitive development by providing opportunities in contexts where integrating diverse sources of knowledge is necessary. However, one of the challenges of its application involves recognizing the adaptation of these tools to specific needs, given the wide range of possibilities they offer to add value to processes associated with the efficient use of natural resources. Furthermore, it highlights the opportunity to implement actions in the short and medium term that are not merely conceptual or indirect, allowing biodiversity to be seen as an enabling factor for the country.

\begin{figure}[h]
    \centering
    \includegraphics[width=0.5\textwidth]{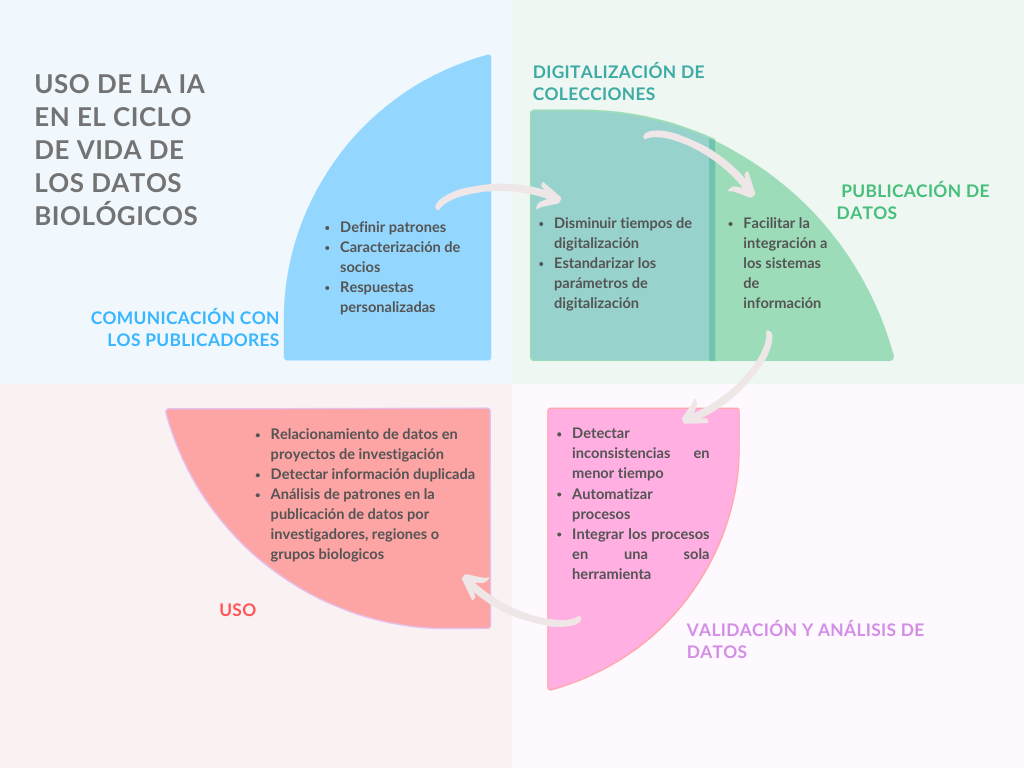}
    \caption{Representación del flujo de los datos biológicos y el uso de la IA en cada etapa. \textit{Figure \ref{fig:Ddata} is translated in the Supplementary material as Figure \ref{fig:data_eng}.}}
    \label{fig:Ddata}
\end{figure}

\subsection*{8. Optimization of Sustainable Use of Colombian Biodiversity: The Role of AI and LangChain in In silico bioprospecting}

\subsubsection*{Context}

Bioprospecting is the systematic search and exploration of biological material and natural sources of molecules and biochemical and genetic information, with the potential to generate valuable products for different sectors \citep{harvey2011}. This process involves not only researching natural resources to discover new compounds, but also cultivating and isolating species that may produce these compounds of interest, followed by experimental validation \citep{kamble2022}. Bioprospecting has been an important focus in Colombia \citep{dnp2018} as it has been recognized since the Convention on Biological Diversity (CBD) as a mechanism to achieve the proposed conservation and use goals of biodiversity \citep{naciones2007}. This initiative is based on the premise that, from the knowledge of biodiversity, it is possible to generate commercially valuable and sustainable products that contribute to the transition toward more sustainable models, such as the bioeconomy, and the consolidation of green businesses and local economies. Although the process provides a mechanism to utilize biodiversity, the conventional approach is often considered inefficient in terms of cost and time, as it depends, among other factors, on the purity and quality of the samples \citep{kamble2022}. This is where new approaches such as in silico bioprospecting arise, using computational tools to accelerate the research process related to the search for bioactive components present in the biome. In silico bioprospecting involves identification of new candidates for product generation using databases and computational tools for the virtual screening of biological information. This method is considered simpler, more effective, and less costly, typically involving two steps: first, exploring databases; and second, in silico characterization, which involves selecting, analyzing, and pre-selecting new candidates using bioinformatics tools.

\subsubsection*{Proposed solution}
Applying in silico bioprospecting or virtual screening \citep{santana2021} as an AI approach that reduces monetary investment and efforts made with traditional research would be a starting point for a better understanding of biological data and its projection within a sustainability context. Specifically, AI allows for the use of preexisting data generated by conventional bioprospecting and leads to the exploration of that information to predict families of metabolites with relevant biological functions. In particular, AI requires complementary tools in data analysis to address research questions before conducting field analyses. In this context, LangChain emerges as a crucial tool by providing models and abstractions to streamline the identification of species with promising potential, ultimately guiding and accelerating the traditional bioprospecting process by focusing efforts and enabling informed decision-making. The objective of this approach is to develop a foundational methodology using these approaches to make the selection route for native Colombian species with bioeconomic potential more efficient, considering that the integration of computational tools and biological data can improve information gathering. Furthermore, it enables the identification of valuable bioactive compounds applicable in various sectors to be validated later in traditional bioprospecting processes.
\subsubsection*{Lessons learned}
There is an opportunity to use Colombia's natural resources sustainably by using AI as an innovative and accessible perspective to improve the country’s bioeconomy. This vision involves integrating AI to support the traditional bioprospecting process, reducing the investment of human and capital resources in these analyses and serving as a promising approach to identify native species with potential based on their biochemical composition. This initiative not only highlights the potential of the tool but also streamlines the selection of species as a mechanism to simultaneously promote human well-being and conservation.
Thus, to strengthen fieldwork, the use of in silico tools is proposed, allowing for imagining and creating new possibilities, dynamics, and relationships while respecting ecological limits established in the context of planetary resilience. Additionally, effective measures are sought to stop the loss of biological diversity while fostering economic alternatives that contribute to and incorporate biodiversity as a key factor to improve human well-being.

\end{paracol}

\begin{table}[h]
\begin{tabular}{|m{8em}|m{34em}|}
\hline
\textbf{Línea estratégica} &
  \textbf{Ejemplo de aplicación de IA} \\ \hline
\multirow{2}{*}{\begin{tabular}[c]{@{}l@{}}Ingredientes \\  naturales\end{tabular}} &
  \begin{tabular}[c]{@{}l@{}}Descubrimiento de compuestos bioactivos lo que puede o no estar asociado \\ con la predicción estructural de las proteínas que lo componen. Puede llegar \\  a ser a nivel molecular, proteico o genómico \citep{ruiz2022}, \\ \citep{nussinov2022}. \end{tabular} \\ \cline{2-2} 
 &
  Modelar y optimizar los modelos de producción de biorrefinerías (para obtención de bioproductos asociados a ingredientes naturales, por ejemplo) ya que la gran cantidad de variables en el esquema en cascada complejizan el proceso \citep{arias2023}. \\ \hline
\multirow{2}{*}{\begin{tabular}[c]{@{}l@{}}Turismo científico \\ de naturaleza\end{tabular}} &
  Modelar, medir y mapear el valor de la naturaleza en el turismo por medio de la cuantificación de la huella espacial o la planificación de acciones de conservación y manejo en el largo plazo (p.ej.: capacidad de carga de la zona) \citep{spalding2023}. \\ \cline{2-2} 
 &
  Turismo inteligente con aplicación en la experiencia turística (p.ej.: actividades y atractivos turísticos), la asistencia virtual, la planificación y ejecución del viaje, el desarrollo industrial, la gestión administrativa, entre muchas otras \citep{li2022}. \\ \hline
\multirow{2}{*}{\begin{tabular}[c]{@{}l@{}}Restauración de \\ ecosistemas\end{tabular}} &
  Identificación y mapeo automático de especies y fuentes semilleras con potencial de restauración en bosques heterogéneos a gran escala (i.e. base para inventarios y acciones de manejo específicas) \citep{beloiu2023}. \\ \cline{2-2} 
 &
  Predicción de germinación de semillas basadas en imágenes digitales en color (RGB) así como el monitoreo de la contaminación, uniformidad física y pureza de las mismas.  \citep{nehoshtan2021}. \\ \hline
\end{tabular}
\caption{Ejemplos de aplicación de IA en el marco del uso y aprovechamiento sostenible de la biodiversidad y sus servicios ecosistémicos en cuatro líneas estratégicas para Colombia. \textit{Table \ref{fig:main_figure} is translated in the Supplementary material as Table \ref{tab:tabla_eng}.}
}
\label{tab:tabla_esp}
\end{table}

\begin{paracol}{2}

\section*{Desafíos y perspectivas}

Nuestras experiencias muestran claramente que la implementación de la inteligencia artificial abre multiplicidad de oportunidades para promover la conservación y uso sostenible de la biodiversidad en Colombia de manera particular y en los trópicos a nivel global. Sin embargo, es necesario tener un entendimiento de las limitaciones y los alcances de la IA para promover una utilización adecuada de esta tecnología. A partir de los casos de estudio descritos anteriormente, destacamos desafíos y oportunidades para implementar exitosamente la IA en labores de conservación y manejo sostenible de recursos biológicos en el Neotrópico.

\subsection*{1. Promover la recolección y gestión soberana de datos de biodiversidad}

La disponibilidad de datos representativos y de calidad es el pilar fundamental para el desarrollo de la tecnología basada en IA en aspectos claves como el entrenamiento de modelos. Mientras que en zonas templadas los registros de biodiversidad son más abundantes y detallados debido a una mayor inversión en investigación, en el Neotrópico, los desafíos logísticos, limitaciones financieras y la complejidad de los ecosistemas han hecho que los datos de biodiversidad estén subrepresentados \citep{collen2008}y con amplias regiones con vacíos de información. Estos datos son las bases de muchos desarrollos en IA, por lo tanto, esta subrepresentación de datos limita la capacidad de los modelos para generalizar y hacer predicciones precisas en esta región. Esto puede generar sesgos en las predicciones, lo que afecta la toma de decisiones y reduce la eficacia de las estrategias de conservación y manejo ambiental \citep{chapman2024} Como resultado, las políticas y acciones implementadas podrían no ajustarse adecuadamente a las condiciones reales del entorno.

Para mejorar la representatividad de los conjuntos de datos sobre biodiversidad, es fundamental seguir con tareas de recolección y catalogación de especies de la mano de expertos y la inclusión de datos generados a partir de modelos participativos. Las labores de trabajo de campo y la taxonomía clásica siguen siendo esenciales para identificar y describir la inexplorada diversidad biológica tropical \citep{mora2011}. Al mismo tiempo, el uso de nuevas tecnologías, como la secuenciación de ADN, sensores remotos, grabadoras de audio y cámaras trampa, puede complementar y agilizar estos procesos, mejorando la calidad y el alcance de los datos recopilados. Además, los datos colectados deben ser sistematizados para desarrollar bases de datos curadas de alta calidad, accesibles y compatibles, con documentación detallada para facilitar su reutilización y validación. Los principios FAIR (Findability, Accessibility, Interoperability, and Reuse) ofrecen un marco para garantizar la transparencia y la ciencia abierta \citep{wilkinson2016}, promoviendo redes colaborativas de generación y uso de conocimiento \citep{vicente2018}. Estos principios FAIR deben ser complementados con los principios CARE (Collective benefit, Authority to control, Responsibility, Ethics). Mientras que los principios FAIR se centran en maximizar el uso y la reutilización de los datos desde una perspectiva técnica y científica, los principios CARE subrayan la importancia de que los datos beneficien a las comunidades locales, respeten su autoridad sobre el acceso y uso de su información, promuevan la responsabilidad social y garanticen prácticas éticamente sólidas \citep{carroll2021}. El uso de la inteligencia artificial apropiadamente implementada puede garantizar la trazabilidad de productos de la biodiversidad lo que en últimas puede disminuir la brecha de distribución de beneficios para comunidades locales e indígenas (Kumming Montreal Agreement 2022).

Las colaboraciones con grandes repositorios de datos del Norte Global han permitido avances en la gestión adecuada de la información. Estas alianzas deben mantenerse y expandirse, pero manteniendo la soberanía de los repositorios gestionados por entidades locales y regionales y garantizando la generación de repositorios espejos en zonas tropicales para salvaguardar la información en el contexto actual de cambios radicales de alta polarización. Esta estrategia fomenta relaciones horizontales en las que múltiples perspectivas son consideradas equitativamente en el diseño y desarrollo de proyectos de IA. Además, permite el desarrollo de nuevos modelos específicamente adaptados al contexto tropical.

\subsection*{2. Desarrollar herramientas para la representación única de la biodiversidad neotropical y sus contextos sociales}

El trópico es uno de los sistemas más complejos a nivel ecológico, lo cual presenta dificultades particulares para el desarrollo y uso de herramientas de IA. La alta biodiversidad tropical, junto con la presencia simultánea de múltiples especies y altos niveles de endemismo, dificultan la identificación precisa de señales específicas en imágenes, sonidos u otros datos. Por ende, exige modelos de IA capaces de discriminar entre muchas clases con datos limitados, adaptándose a la especialización y microhábitats específicos para optimizar la detección y seguimiento de especies \citep{van2017}. La implementación de modelos globales sin evaluaciones contextuales puede generar interpretaciones erróneas y promover una dependencia tecnológica que ignora las características únicas de cada región.

La opción más sencilla para adaptar los modelos de IA al contexto tropical, consiste en evaluar los sesgos existentes en los modelos. Por ejemplo, se ha demostrado que los modelos usados para identificar especies en aplicaciones como BirdNET tienen una mayor precisión en especies de regiones con muchos registros (como EE. UU. y Europa), mientras que presentan errores frecuentes en especies tropicales con menos datos de entrenamiento \citep{de2024}. Estos sesgos pueden llevar a falsas ausencias o subestimaciones de especies vulnerables, afectando las estrategias de conservación. Para minimizar estos sesgos, se pueden implementar métricas que midan sesgos geográficos y taxonómicos en los modelos, permitiendo su calibración y ajuste de predicciones con base en estos análisis \citep{wood2024}.

Si se cuenta con más experiencia en ML, una aproximación efectiva es el aprendizaje por transferencia, o transfer learning. Esta es una técnica de aprendizaje auaseático que permite ajustar modelos pre-entrenados a nuevos contextos, reduciendo la necesidad de grandes volúmenes de datos de entrenamiento. En el transfer learning, se conserva la capacidad de un modelo para discriminar patrones generales (capturados en las primeras capas de una red neuronal), pero se ajustan las capas finales para adaptar esta capacidad a nuevos entornos, mejorando su desempeño en nuevas tareas. Esta técnica facilita adaptar modelos a las condiciones ambientales y ecológicas específicas de los ecosistemas tropicales, y en general en proyectos donde los datos de entrenamiento son escasos. Además, esta técnica minimiza los requerimientos en infraestructura tecnológica, lo cual presenta una ventaja dada la limitada capacidad en muchas zonas del trópico. También se pueden considerar técnicas de aprendizaje semi-supervisado, los cuales utilizan una combinación de datos etiquetados y no etiquetados para mejorar la capacidad de reconocimiento de clases raras \citep{ma2024}. La aplicación de estas técnicas de ML permitirá desarrollar modelos adaptados a cada región, reduciendo la dependencia de tecnologías externas y fomentando nuevas colaboraciones regionales.

Finalmente, es necesario considerar que nuestra región cuenta con contextos sociales únicos, como su gran diversidad cultural y lingüística. La pluralidad de culturas y lenguas en las regiones tropicales exige el desarrollo de herramientas de IA con interfaces y documentación adaptadas a distintos contextos culturales, asegurando su accesibilidad y utilidad para diversas comunidades. Además, la gestión comunitaria y el conocimiento tradicional desempeñan un papel fundamental en la conservación. La participación activa de comunidades locales e indígenas, que poseen un conocimiento profundo del entorno, es crucial para la recolección y validación de datos, lo que permite desarrollar modelos más representativos y precisos. En general, para lograr que las herramientas tengan un impacto real en la conservación, es necesario integrar los conflictos socioambientales específicos de la región, como la expansión de la frontera agrícola, la minería ilegal o la deforestación, en el modelamiento para la gestión de recursos naturales y el manejo territorial. Esto implica la incorporación de perspectivas de múltiples actores—comunidades locales, tomadores de decisiones, científicos y organizaciones ambientales—para apoyar estrategias de conservación más equitativas y efectivas.

\subsection*{3. Integrar capacidades humanas y garantizar la transparencia en sistemas de IA}

La implementación exitosa de la automatización a través de la IA no ocurre instantáneamente, sino que implica un proceso gradual de mejoras continuas. Este proceso involucra distintos tipos de interacciones: entre sistemas de IA, con el entorno y a través del autoaprendizaje. Sin embargo, la interacción humana sigue siendo la más fundamental, ya que es la que orienta, supervisa y adapta las herramientas para que respondan a nuestros objetivos y necesidades. La optimización constante y la retroalimentación, impulsadas por la participación humana, garantizan que la automatización no solo sea eficiente, sino también alineada con valores y propósitos específicos \citep{stenhouse2023}.

Integrar los sistemas de IA en un entorno de aprendizaje con intervención experta no solo permite evaluar el desempeño, sino también recopilar nuevas muestras que ayuden a perfeccionar y ajustar los modelos. Esta aproximación conocida como aprendizaje activo, optimiza el uso de datos, maximiza el tiempo de los expertos y mejora el desempeño de los modelos Este enfoque resulta particularmente relevante en contextos con lagunas de información, como en el trópico, donde el conocimiento de comunidades locales y de expertos en campo es crucial para identificar especies y patrones ecológicos poco estudiados \citep{wearn2019}.

La IA sólo puede potenciar el conocimiento local si se desarrolla con supervisión humana. La responsabilidad en el uso de esta tecnología debe recaer en las personas para asegurar que su implementación se alinee con valores, objetivos y principios éticos. Esto garantiza que la implementación de la IA considere el contexto local, adaptando los modelos a realidades ecológicas, culturales y sociales específicas, asegurando que la automatización no sustituya el conocimiento local, sino que lo potencie. No obstante, la complejidad de los modelos actuales, muchas veces percibidos como cajas negras, dificulta su adopción al generar desconfianza y limitar su interpretación \citep{lucas2020}. 
Promover la transparencia en el desarrollo y uso de los modelos de IA facilita su comprensión, fomenta la confianza de los usuarios y permite una adopción más responsable \citep{wearn2019}. Generar transparencia implica implementar estrategias que hagan interpretables estos modelos, permitiendo identificar y mitigar sesgos o impactos no deseados \citep{doshi2017, rudin2019}. De este modo, se garantiza un uso más ético y alineado con principios normativos, además de mejorar la supervisión y toma de decisiones. Esto asegura que la IA sea una herramienta confiable y adaptable a diferentes contextos y necesidades.

La integración de la IA debe ser un equilibrio entre innovación tecnológica y supervisión humana. Solo a través de la transparencia, la adaptabilidad y la responsabilidad en su uso, la IA podrá convertirse en una herramienta confiable y efectiva que complemente el conocimiento local y contribuya a la toma de decisiones informadas. 

\subsection*{4. Empoderar a las comunidades a través de datos y modelos abiertos de calidad}

Las plataformas de ciencia participativa han transformado la forma en que las personas se conectan con la biodiversidad, permitiendo la recopilación de datos a una escala sin precedentes. Por ejemplo, iNaturalist permite a los usuarios identificar especies en imágenes mediante IA, mientras que Merlin Bird ID permite a los usuarios identificar aves mediante imágenes, descripciones y grabaciones de audio. El desarrollo de estas plataformas sigue un círculo virtuoso: los usuarios cuentan con una plataforma intuitiva para registrar y seguir sus observaciones de fauna o flora. Así, contribuyen con grandes volúmenes de datos, como imágenes, grabaciones o videos, los cuales son utilizados para entrenar modelos que mejoran continuamente la identificación de especies. Estos avances no solo facilitan nuevos registros, sino que también incentivan la participación de más personas en la colecta de datos. Este proceso, impulsado por la disponibilidad de datos abiertos, fomenta la generación y democratización del conocimiento sobre la biodiversidad en general, pero sobre todo la biodiversidad local.

En este contexto, plataformas de ciencia participativa con modelos de IA pueden servir como una poderosa interfaz para facilitar compartir y buscar información, fomentando la creación de comunidad y fortaleciendo la conexión de las personas con el mundo natural que las rodea. Comprender y valorar el entorno natural es esencial para generar acciones de conservación, ya que la protección de la biodiversidad está intrínsecamente vinculada al grado de conocimiento y aprecio que las personas tienen por ella \citep{hunter2003}. 
Así como los datos, los modelos abiertos son el complemento para democratizar el acceso a herramientas científicas, especialmente en regiones con recursos limitados. Al eliminar barreras económicas y tecnológicas, permiten que más investigadores, instituciones y comunidades locales utilicen y desarrollen estas herramientas. Además, fomentan la innovación y aceleran el progreso científico, ya que cualquier experto puede mejorar, adaptar y compartir sus avances, generando soluciones más eficientes y ajustadas a cada contexto. Asimismo, promueven la transparencia y reproducibilidad en la ciencia, facilitando la adaptación tecnológica a las necesidades locales y fortaleciendo la toma de decisiones informadas en biodiversidad.

De forma general, los datos, modelos y la ciencia abierta fomentan el desarrollo de tecnología en la región. Sin embargo, existen aún diferentes barreras a esta apertura de datos y conocimiento: i) la falta de atribución correcta a las personas y organizaciones implicadas en las investigaciones; ii) el manejo de datos sensibles que deben tener ciertas restricciones (p.ej.: la localización de especies amenazadas o vulnerables al tráfico ilegal); iii) la falta de recursos para hacer una correcta gestión de los datos (p.ej.: almacenaje, curaduría, anotación, estandarización, sistematización, publicación, etc.); iv) ausencia de infraestructura y financiación sostenida en el tiempo para custodiar la información a largo plazo y mantenerla actualizada \citep{paic2021}. Es entonces fundamental continuar con el esfuerzo por seguir los principios FAIR y CARE, mencionados en anteriormente, con el fin de superar las barreras antes descritas a la hora de operacionalizar estos principios.  Estos principios se alinean con el artículo 6 del Acuerdo de Escazú, el cual exige que los Estados deben generar, compilar, sistematizar y publicar información ambiental. Además resalta que esta información sea reutilizable, procesable y esté disponible en formatos accesibles, asegurando que llegue a todas las comunidades, incluidas las más vulnerables \citep{comision22}. Superar estos desafíos requiere un compromiso continuo con la ciencia abierta, impulsando políticas, financiamiento e infraestructura que garanticen una gestión de datos ética, accesible y sostenible, fortaleciendo así el desarrollo tecnológico y la conservación de la biodiversidad en la región.

\subsection*{5. Fomentar colaboraciones entre las comunidades de conservación de la biodiversidad y computación}

Actualmente, existe una brecha significativa entre las ciencias biológicas y las ciencias de la computación. Por un lado, los ecólogos y biólogos enfrentan preguntas complejas sobre los ecosistemas neotropicales, pero en algunos casos tienen acceso limitado a herramientas computacionales avanzadas. Por otro lado, los científicos de la computación poseen un alto nivel técnico en analítica de datos, pero carecen del contexto ecológico necesario para adaptar sus herramientas a problemas específicos de conservación. Reducir esta brecha es esencial para desarrollar soluciones tecnológicas más efectivas.

Las colaboraciones interdisciplinarias pueden aprovechar tanto el conocimiento ecológico local como las técnicas computacionales avanzadas para abordar desafíos clave en conservación y gestión de recursos naturales. Desde el ámbito de la ecología, investigadoras y profesionales pueden aportar datos curados y estructurados, además de definir con claridad los problemas que necesitan resolverse. Estos datos sirven como base para entrenar modelos de aprendizaje automático, los cuales dependen de características bien definidas y etiquetas precisas para hacer predicciones y clasificaciones efectivas. Garantizar la calidad y representatividad de estos conjuntos de datos es fundamental para que los modelos puedan generalizar adecuadamente a nuevas situaciones.

Desde las ciencias de la computación, es crucial desarrollar herramientas accesibles y de código abierto que faciliten su adopción por parte de la comunidad ecológica. Además, la capacitación juega un papel clave: ofrecer cursos, guías y talleres sobre el desarrollo y evaluación de modelos de IA puede empoderar a los ecólogos para integrar estas tecnologías en su trabajo \citep{cole2023}. Iniciativas como Latin American Meeting
in Artificial Intelligence \citep{khipu} y LatinX in AI \citep{latinX}, que promueven la capacitación en IA en América Latina, representan espacios estratégicos para fomentar estas colaboraciones interdisciplinarias.

Finalmente, es fundamental fomentar un pensamiento crítico y auto-reflexivo sobre las capacidades y limitaciones de las herramientas de IA. A medida que emergen nuevas tecnologías, su uso en ecología y conservación debe basarse en una comprensión clara de sus alcances y restricciones, evitando interpretaciones erróneas o aplicaciones inadecuadas. Es importante además que se empiezan a evaluar que las intervenciones implementadas a partir de soluciones generados con IA en temas complejos como la conservación de la biodiversidad, esto puede realimentar de manera virtuosa los usos y aproximaciones aprendidas.

\subsection*{6. Construir alianzas globales equitativas}

Los valores y contextos epistémicos en los que han crecido los estudios de la biodiversidad han estado acompañados de prácticas coloniales y extractivistas \citep{trisos2021}. Por ejemplo, la ciencia helicóptero, donde investigadoras del Norte Global extraen datos del Sur Global sin ninguna relación con el contexto socio-ecológico y científico, ha sido una práctica común que hoy en día continua. 

Las estructuras de dominación tienen distintos niveles que llevan a relaciones de poder asimétricas. Un ejemplo ocurre en las colaboraciones Norte-Sur, en las cuales la influencia de los investigadores del Norte Global en las decisiones metodológicas se impone debido al prestigio de universidades, acceso a financiación, incentivos de investigación, entre otros. Esta no es solo una práctica que se da en contextos de colaboraciones Norte-Sur; aún entre investigadores del Sur Global existe un colonialismo interno, dado por jerarquías de poder \citep{rivera2012, hirschfeld2023}. Un caso común es la hegemonía cultural de una disciplina sobre otra o entre lo cuantitativo sobre lo cualitativo. Por ejemplo, cuando no se reconoce ni se da crédito a la investigación en los procesos de recolección y curación de datos o excluyendo el papel de las comunidades como generadores de conocimiento.

Estas prácticas pueden continuar en el desarrollo de modelos de IA para la Biodiversidad, donde solo se ve al Neotrópico como ese lugar exótico en el que sucede el feńomeno de estudio, donde se extraen los datos para ser desarrollados en el Norte Global, pero no como un lugar donde se puede pensar y desarrollar también el fenómeno de estudio. Potenciales consecuencias negativas de esto es la monocultura científica de la agenda de investigación \citep{messeri2024artificial} y aún más , la imposición de agendas de conservación que no incluyan las visiones y experiencias del sur Global. 

Es urgente integrar de manera integral las particularidades de las culturas latinoamericanas en la creación de tecnologías de IA para la región; es decir, desarrollar tecnologías pensadas para y con los latinoamericanos, valorando su participación en investigación y desarrollo, en lugar de limitarlos a ser meros proveedores de datos o anotaciones manuales con bajo valor agregado. Dentro de estas colaboraciones se debe incluir tácticas de previsión, tales como, transferencia de conocimiento, atribución de créditos, colaboraciones a largo plazo, incluir a comunidades y su contexto, reconocer las teorías decoloniales del Sur Global en los proyectos,  analizar las formas de producción de conocimiento y construir modelos utilizando datos propios de cada región en colaboraciones horizontales \citep{armenteras2021, mohamed2020, haelewaters2021}. Solo a través de alianzas equitativas y horizontales, que reconozcan y valoren las voces, conocimientos y necesidades del Sur Global, será posible desarrollar tecnologías de IA para la biodiversidad que sean verdaderamente inclusivas, sostenibles y alineadas con los contextos socioecológicos de la región.

\switchcolumn

\section*{Challenges and perspectives}

Our experiences clearly show that the implementation of artificial intelligence opens up numerous opportunities to promote the conservation and sustainable use of biodiversity in Colombia and especially in the tropics. However, it is necessary to understand the limitations and scope of AI in order to promote its proper use. Based on the case studies described above, we highlight challenges and opportunities for successfully implementing AI in conservation efforts and sustainable management of biological resources in the Neotropics.
\\

\subsection*{1. Promote the sovereign collection and management of biodiversity data}

The availability of representative and high-quality data is the cornerstone for the development of AI-based technology, particularly in key aspects such as model training. While biodiversity records in temperate zones are more abundant and detailed due to greater investment in research, in the Neotropics, logistical challenges, financial constraints, and the complexity of ecosystems have led to underrepresented biodiversity data \citep{collen2008}, with vast regions experiencing data gaps. These data form the foundation of many AI developments; therefore, this underrepresentation limits the ability of models to generalize and make accurate predictions in this region. This can introduce biases in predictions, affecting decision-making and reducing the effectiveness of conservation and environmental management strategies \citep{chapman2024}. As a result, policies and actions implemented may not adequately align with the actual conditions of the environment.

To improve the representativeness of biodiversity datasets, it is essential to continue species collection and cataloging efforts in collaboration with experts, as well as to incorporate data generated through participatory models. Fieldwork and classical taxonomy remain crucial for identifying and describing the unexplored biological diversity of the tropics \citep{mora2011}. At the same time, the use of new technologies, such as DNA sequencing, remote sensors, audio recorders, and camera traps, can complement and expedite these processes, enhancing the quality and scope of collected data. Additionally, the data collected must be systematized to develop high-quality, curated databases that are accessible, interoperable, and well-documented to facilitate reuse and validation. The FAIR principles (Findability, Accessibility, Interoperability, and Reuse) provide a framework to ensure transparency and open science \citep{wilkinson2016}, fostering collaborative networks for knowledge generation and use \citep{vicente2018}. These FAIR principles should be complemented by the CARE principles (Collective Benefit, Authority to Control, Responsibility, Ethics). While FAIR principles focus on maximizing the use and reuse of data from a technical and scientific perspective, CARE principles emphasize the importance of ensuring that data benefit local communities, respect their authority over access and use of their information, promote social responsibility, and uphold ethical practices \citep{carroll2021}. Appropriately implemented artificial intelligence can ensure the traceability of biodiversity products, ultimately helping to reduce the benefit-sharing gap for local and Indigenous communities (Kunming-Montreal Agreement 2022).

Collaborations with large data repositories in the Global North have enabled progress in the effective management of information. These partnerships should be maintained and expanded, but with a focus on preserving the sovereignty of repositories managed by local and regional entities. Additionally, the creation of mirror repositories in tropical regions is essential to safeguard information in the current context of radical and highly polarized changes. This strategy fosters horizontal relationships in which multiple perspectives are equitably considered in the design and development of AI projects. Furthermore, it enables the development of new models specifically adapted to the tropical context.

\subsection*{2. Develop tools for the unique representation of Neotropical biodiversity and its social contexts}

The tropics represent one of the most ecologically complex systems, posing particular challenges for the development and application of AI tools. The high biodiversity, the simultaneous presence of multiple species, and elevated levels of endemism make it difficult to accurately identify specific signals in images, sounds, or other data. Consequently, AI models must be capable of distinguishing among numerous classes with limited data, adapting to species specialization and microhabitats to optimize detection and monitoring \citep{van2017}. The implementation of global models without contextual evaluations may lead to misinterpretations and foster technological dependencies that overlook the unique characteristics of each region.

The simplest approach to adapting AI models to tropical contexts is to assess existing biases within these models. For instance, studies have shown that species-identification models used in applications such as BirdNET achieve higher accuracy for species in well-documented regions (such as the U.S. and Europe) but frequently misidentify tropical species with fewer training data \citep{de2024}. These biases can lead to false absences or underestimations of vulnerable species, impacting conservation strategies. To mitigate these biases, geographic and taxonomic bias metrics can be implemented, allowing for model calibration and prediction adjustments based on these analyses \citep{wood2024}.

For those with more experience in machine learning (ML), transfer learning is an effective approach. This technique enables the adaptation of pre-trained models to new contexts, reducing the need for large volumes of training data. In transfer learning, a model retains its ability to discriminate general patterns (captured in the initial layers of a neural network) while the final layers are fine-tuned to adjust this capacity to new environments, improving performance on novel tasks. This technique facilitates the adaptation of models to the specific environmental and ecological conditions of tropical ecosystems, and it is particularly useful in projects where training data are scarce. Additionally, transfer learning reduces technological infrastructure requirements, which is advantageous given the limited capacity in many tropical regions. Semi-supervised learning techniques can also be considered, as they leverage a combination of labeled and unlabeled data to enhance the recognition of rare classes \citep{ma2024}. The application of these ML techniques will allow for the development of regionally adapted models, reducing dependency on external technologies and fostering new regional collaborations.

Finally, it is essential to acknowledge the unique social contexts of tropical regions, including their vast cultural and linguistic diversity. The plurality of cultures and languages demands the development of AI tools with interfaces and documentation tailored to different cultural contexts, ensuring accessibility and usability for diverse communities. Moreover, community-based management and traditional knowledge play a fundamental role in conservation. The active participation of local and Indigenous communities, who possess deep ecological knowledge, is crucial for data collection and validation, leading to the development of more representative and accurate models. In general, to ensure that AI tools have a meaningful impact on conservation, it is necessary to integrate the region's specific socio-environmental conflicts—such as agricultural expansion, illegal mining, and deforestation—into modeling efforts for natural resource management and land-use planning. This requires incorporating the perspectives of multiple stakeholders—local communities, decision-makers, scientists, and environmental organizations—to support more equitable and effective conservation strategies.
\\

\subsection*{3. Integrate human capabilities and ensure transparency in AI system}

The successful implementation of automation through AI does not occur instantaneously, but rather involves a gradual process of continuous improvements. This process encompasses various types of interactions: among AI systems, with the environment, and through self-learning. However, human interaction remains the most fundamental, as it guides, supervises, and adapts the tools to meet our objectives and needs. Constant optimization and feedback, driven by human participation, ensure that automation is not only efficient but also aligned with specific values and purposes \citep{stenhouse2023}.

Integrating AI systems into a learning environment with expert intervention not only allows for performance evaluation but also facilitates the collection of new samples that aid in refining and adjusting the models. This approach, known as active learning, optimizes data usage, maximizes expert time, and enhances model performance. This approach is particularly relevant in contexts with information gaps, such as in the tropics, where knowledge from local communities and field experts is crucial for identifying understudied species and ecological patterns \citep{wearn2019}.

AI can only enhance local knowledge if it is developed with human oversight. The responsibility for the use of this technology must rest with people to ensure that its implementation aligns with ethical values, objectives, and principles. This ensures that the implementation of AI considers the local context, adapting models to specific ecological, cultural, and social realities, ensuring that automation does not replace local knowledge but rather enhances it. Nevertheless, the complexity of current models, often perceived as black boxes, hinders their adoption by generating distrust and limiting their interpretation \citep{lucas2020}.

Promoting transparency in the development and use of AI models facilitates their understanding, fosters user trust, and enables more responsible adoption \citep{wearn2019}. Generating transparency involves implementing strategies that make these models interpretable, allowing for the identification and mitigation of biases or unintended impacts \citep{doshi2017, rudin2019}. This ensures a more ethical use aligned with normative principles, in addition to improving supervision and decision-making. This ensures that AI is a reliable and adaptable tool to different contexts and needs.

The integration of AI must be a balance between technological innovation and human oversight. Only through transparency, adaptability, and responsibility in its use can AI become a reliable and effective tool that complements local knowledge and contributes to informed decision-making.

\subsection*{4. Empowering communities through quality open data and models}

Participatory science platforms have transformed how people engage with biodiversity, enabling data collection on an unprecedented scale. For example, iNaturalist allows users to identify species in images using AI, while Merlin Bird ID enables users to identify birds through images, descriptions, and audio recordings. The development of these platforms follows a virtuous cycle: users are provided with an intuitive platform to record and track their observations of flora and fauna. In doing so, they contribute large volumes of data, such as images, recordings, or videos, which are used to train models that continuously improve species identification. These advancements not only facilitate new records but also encourage greater public participation in data collection. This process, driven by the availability of open data, fosters the generation and democratization of knowledge about biodiversity in general, and particularly local biodiversity.
\\
\\

In this context, participatory science platforms with AI models can serve as powerful interfaces to facilitate information sharing and retrieval, fostering community building and strengthening people's connection to the natural world around them. Understanding and valuing the natural environment is essential for driving conservation actions, as the protection of biodiversity is intrinsically linked to the level of knowledge and appreciation people have for it \citep{hunter2003}.

Just as open data is crucial, open models are the complement needed to democratize access to scientific tools, especially in regions with limited resources. By removing economic and technological barriers, they enable more researchers, institutions, and local communities to use and develop these tools. Furthermore, they promote innovation and accelerate scientific progress, as any expert can improve, adapt, and share their advancements, generating more efficient and context-specific solutions. They also promote transparency and reproducibility in science, facilitating technological adaptation to local needs and strengthening informed decision-making in biodiversity.

Overall, open data, models, and open science foster technological development in the region. However, several barriers to this openness of data and knowledge remain: i) the lack of proper attribution to individuals and organizations involved in research; ii) the handling of sensitive data that requires certain restrictions (e.g., the location of species threatened by or vulnerable to illegal trafficking); iii) the lack of resources for proper data management (e.g., storage, curation, annotation, standardization, systematization, publication, etc.); and iv) the absence of sustained infrastructure and funding to safeguard information over the long term and keep it updated \citep{paic2021}. It is therefore essential to continue efforts to adhere to the FAIR and CARE principles, mentioned earlier, to overcome these barriers when operationalizing these principles. These principles align with Article 6 of the Escazú Agreement, which requires states to generate, compile, systematize, and publish environmental information. It also emphasizes that this information must be reusable, processable, and available in accessible formats, ensuring it reaches all communities, including the most vulnerable \citep{comision22}. Overcoming these challenges requires a sustained commitment to open science, driving policies, funding, and infrastructure that ensure ethical, accessible, and sustainable data management, thereby strengthening technological development and biodiversity conservation in the region.

\subsection*{5. Foster collaborations between biodiversity conservation communities and computing}

Currently, there is a significant gap between the biological sciences and computer science. On one hand, ecologists and biologists face complex questions regarding Neotropical ecosystems, yet they often have limited access to advanced computational tools. On the other hand, computer scientists possess high technical expertise in data analytics but often lack the ecological context necessary to adapt their tools to specific conservation challenges. Bridging this gap is essential for developing more effective technological solutions.

Interdisciplinary collaborations can leverage both local ecological knowledge and advanced computational techniques to address key challenges in conservation and natural resource management. From an ecological perspective, researchers and professionals can contribute curated and structured datasets while clearly defining the problems that need to be solved. These datasets serve as the foundation for training machine learning models, which rely on well-defined features and accurate labels to make effective predictions and classifications. Ensuring the quality and representativeness of these datasets is crucial for models to generalize adequately to new scenarios.

From a computer science perspective, it is critical to develop accessible and open-source tools that facilitate adoption by the ecological community. Additionally, training plays a key role: offering courses, guides, and workshops on AI model development and evaluation can empower ecologists to integrate these technologies into their work \citep{cole2023}. Initiatives such as the Latin American Meeting in Artificial Intelligence Khipu \citep{khipu} and LatinX in AI (LXAI) \citep{latinX}, which promote AI training in Latin America, provide strategic spaces to foster interdisciplinary collaborations.

Finally, it is essential to encourage critical and self-reflective thinking regarding the capabilities and limitations of AI tools. As new technologies emerge, their application in ecology and conservation should be grounded in a clear understanding of their scope and constraints to avoid misinterpretations or inappropriate uses. Moreover, it is crucial to assess the interventions implemented based on AI-generated solutions in complex areas such as biodiversity conservation. This process can create a virtuous feedback loop, refining approaches and improving future applications.

\subsection*{6. Build equitable global alliances}

The epistemic values and contexts in which biodiversity studies have developed have been accompanied by colonial and extractivist practices \citep{trisos2021}. For example, helicopter science, where researchers from the Global North extract data from the Global South without any engagement with the socio-ecological and scientific context, has been a common practice that persists today.

Structures of domination have various levels that lead to asymmetric power relations. An example occurs in North-South collaborations, in which the influence of Global North researchers in methodological decisions is imposed due to the prestige of universities, access to funding, research incentives, among others. This is not only a practice that occurs in North-South collaboration contexts; even among researchers from the Global South, internal colonialism exists, driven by power hierarchies \citep{rivera2012, hirschfeld2023}. A common case is the cultural hegemony of one discipline over another or of the quantitative over the qualitative. For example, when research in data collection and curation processes is not recognized or credited, or when the role of communities as knowledge generators is excluded.

These practices can continue in the development of AI models for Biodiversity, where the Neotropics are seen only as an exotic place where the phenomenon of study occurs, where data are extracted to be developed in the Global North, but not as a place where the phenomenon of study can also be conceived and developed. Potential negative consequences of this include the scientific monoculture of the research agenda \citep{messeri2024artificial} and, even more so, the imposition of conservation agendas that do not include the visions and experiences of the Global South.

It is urgent to comprehensively integrate the particularities of Latin American cultures in the creation of AI technologies for the region; that is, to develop technologies designed for and with Latin Americans, valuing their participation in research and development, rather than limiting them to being mere providers of data or manual annotations with low added value. Within these collaborations, foresight tactics should be included, such as knowledge transfer, credit attribution, long-term collaborations, inclusion of communities and their context, recognition of decolonial theories from the Global South in projects, analysis of knowledge production forms, and construction of models using data from each region in horizontal collaborations \citep{armenteras2021, mohamed2020, haelewaters2021}. Only through equitable and horizontal alliances, which recognize and value the voices, knowledge, and needs of the Global South, will it be possible to develop AI technologies for biodiversity that are truly inclusive, sustainable, and aligned with the socio-ecological contexts of the region.

\end{paracol}

\begin{paracol}{2}

\section*{Conclusiones}

La sinergia entre la IA y  los desafíos que enfrenta la conservación de la biodiversidad representa una oportunidad única para diseñar estrategias innovadoras y sostenibles, siempre que se priorice la equidad, la transparencia y la colaboración interdisciplinaria entre actores involucrados. En los entornos neotropicales, donde la alta diversidad biológica y la complejidad socioeconómica contrastan con los recursos financieros limitados, la IA puede optimizar y escalar procesos de sostenibilidad de manera más eficiente. 

Nuestras experiencias resaltan las grandes oportunidades de la IA en multiplicidad de agendas de investigación sobre biodiversidad. Además, reconocemos que esta tecnología no es una solución única ni automática para los desafíos ambientales de la región. Su impacto está condicionado por el contexto sociopolítico en el que se desarrolla e implementa, lo que implica que su uso puede reforzar desigualdades o, por el contrario, abrir nuevas oportunidades de colaboración y empoderamiento. Por ello, es fundamental evitar un enfoque simplista o excesivamente optimista. La implementación de la IA debe considerar una perspectiva plural que integre las humanidades y las ciencias biológicas, entre otras áreas. Además, el desarrollo y uso de estas herramientas debe ser riguroso, inclusivo y transparente, garantizando que las soluciones tecnológicas respondan a las necesidades reales de conservación. Implementada de manera ética y colaborativa, la IA puede contribuir significativamente a mejorar la recopilación, el análisis y la interpretación de datos sobre biodiversidad. 

Integrar la IA en las políticas y estrategias de biodiversidad no solo puede optimizar la toma de decisiones, sino también fortalecer la capacidad de los países para responder a los desafíos ambientales de manera efectiva. Durante el 2022 se acordó el Marco Mundial Kunming-Montreal de la Diversidad Biológica, el cual definió 23 metas específicas para frenar la pérdida de la biodiversidad. Como consecuencia, los países participantes han empezado a plasmar estos objetivos en metas nacionales y regionales usando los Planes de Acción Nacionales de Biodiversidad (PAB). A la par, durante los últimos años los gobiernos han creado políticas nacionales de IA que han definido sus estrategias, narrativas y visiones para el desarrollo de sistemas de IA \citep{galindo2021, zambrano2021, sanchez2021}. 

Por ejemplo, Colombia renovó su PAB durante octubre del 2024 con 4 apuestas, 6 metas nacionales y 191 acciones \citep{ministerio2024}, mientras que la Política Nacional de IA (PNIA) presentada en febrero del 2025 cuenta con 6 ejes estratégicos y 106 acciones las cuales buscan definir condiciones habilitantes que permitan el aprovechamiento de esta tecnología al enfrentar diversos problemas sociales, económicos y ambientales mientras que también proponen regulaciones para mitigar sus riesgos \citep{dnp2025}; ambas iniciativas planean ejecutarse hasta el 2030.

La convergencia temporal de estos dos documentos de política crea una oportunidad de inversión financiera tanto de en los sectores de IA como de conservación donde se pueden generar proyectos comunes. Por lo tanto, es fundamental que durante los próximos años se puedan coordinar los diversos objetivos y acciones de ambos instrumentos, para que permitan el desarrollo de innovación en IA motivada por su aplicación en biodiversidad, fomentando colaboraciones no sólo entre estas áreas de investigación, sino también en las entidades públicas que se encargan de la ejecución de estos planes y estrategias.
De esta manera, los gobiernos, empresas e instituciones de la región podrán aprovechar los conocimientos basados en evidencia para desarrollar políticas más efectivas que atiendan tanto el bienestar humano como el de las demás especies que habitan el planeta. A su vez, los ciudadanos podrán conocer mejor su biodiversidad, fomentando una mayor conciencia y participación en la conservación de su entorno natural. Solo con una implementación ética y alineada con las necesidades regionales, la IA podrá convertirse en una herramienta clave para fortalecer la conservación de la biodiversidad y el bienestar humano. 

\subsection*{Descargo de responsabilidad}

Las opiniones presentadas en este documento representan las opiniones personales de los autores y no reflejan necesariamente las políticas o posiciones oficiales de sus organizaciones.

\switchcolumn

\section*{Conclusions}

The synergy between AI and the challenges of biodiversity conservation represents a unique opportunity to design innovative and sustainable strategies, always promoting equity, transparency, and interdisciplinary collaboration between stakeholders. In Neotropical environments, where extraordinary biological diversity and socioeconomic complexity contrast with available financial resources, AI can optimize and scale large-scale sustainability processes more efficiently. 

Our experiences highlight the great opportunities of AI in the multiplicity research agendas related to AI. In addition, we acknowledge that this technology is not a one-size-fits-all or automatic solution to environmental challenges in the region. Its impact is conditioned by the sociopolitical context in which it is developed and implemented, meaning that its use can either reinforce inequalities or, conversely, open new opportunities for collaboration and empowerment. Therefore, it is essential to avoid a simplistic or overly optimistic approach. Instead, the implementation of the IA should consider a pluralist perspective that integrates humanities, biological sciences, and other areas. Also, the development and use of these tools should be rigorous, inclusive, and transparent, ensuring that technological solutions align with real conservation needs. If implemented ethically and collaboratively, AI can contribute significantly to improving the collection, analysis, and interpretation of biodiversity data.

Integrating AI into policy and biodiversity strategies can not only optimize decision making, but also strengthen the capacity of countries to respond to environmental challenges more effectively. During 2022, the Kunming-Montreal Global Biodiversity Framework, which defines 23 targets to slow biodiversity loss, was agreed. As a consequence, countries have been starting to capture these goals in national and regional goals using National Biodiversity Strategies and Action Plans (NBSAPs). Concurrent, in recent years governments have been working on national policies around AI to define strategies, narratives, and visions for the development of AI systems \citep{galindo2021, zambrano2021, sanchez2021}. 

For example, Colombia updated its NBSAP in October 2024, which includes 4 strategic priorities, 6 national goals, and 191 actions \citep{ministerio2024}. Meanwhile, the National AI Policy, presented in February 2025, outlines 6 strategic pillars and 106 actions aimed at creating enabling conditions to leverage AI in addressing social, economic, and environmental challenges, while also proposing regulations to mitigate its risks \citep{dnp2025}. Both initiatives are planned to be implemented until 2030.

The temporal convergence of these two policy documents creates a unique opportunity for financial investment in both the AI and conservation sectors, where joint projects can be developed. Therefore, it is crucial that over the coming years, the various objectives and actions of both frameworks are coordinated to foster AI innovation driven by its application in biodiversity. This will encourage collaborations not only between these research areas but also among the public entities responsible for executing these plans and strategies. In this way, governments, businesses, and institutions in the region can leverage evidence-based knowledge to develop more effective policies that address both human well-being and the well-being of other species on the planet. At the same time, citizens will gain a better understanding of their biodiversity, fostering greater awareness and participation in the conservation of their natural environment. Only through ethical implementation aligned with regional needs can AI become a key tool for strengthening biodiversity conservation and human well-being.
\\

\subsection*{Disclaimer}

The opinions presented in this document represent the personal views of the authors and do not necessarily reflect the policies or official positions of their organizations.
    
\end{paracol}

\bibliographystyle{abbrvnat}
\bibliography{references}

\begin{thebibliography}{106}
\providecommand{\natexlab}[1]{#1}
\providecommand{\url}[1]{\texttt{#1}}
\expandafter\ifx\csname urlstyle\endcsname\relax
  \providecommand{\doi}[1]{doi: #1}\else
  \providecommand{\doi}{doi: \begingroup \urlstyle{rm}\Url}\fi

\bibitem[Ahumada et~al.(2020)Ahumada, Fegraus, Birch, Flores, Kays, O’Brien, Palmer, Schuttler, Zhao, Jetz, et~al.]{ahumada2020wildlife}
J.~A. Ahumada, E.~Fegraus, T.~Birch, N.~Flores, R.~Kays, T.~G. O’Brien, J.~Palmer, S.~Schuttler, J.~Y. Zhao, W.~Jetz, et~al.
\newblock Wildlife insights: A platform to maximize the potential of camera trap and other passive sensor wildlife data for the planet.
\newblock \emph{Environmental Conservation}, 47\penalty0 (1):\penalty0 1--6, 2020.

\bibitem[Allard et~al.(2023)Allard, Webber, Sundberg, and Brown]{allard2023}
A.~Allard, L.~Webber, J.~H. Sundberg, and A.~Brown.
\newblock New and changing use of technologies in monitoring: Drones, artificial intelligence, and environmental dna.
\newblock In A.~Allard, E.~C.~H. Keskitalo, and A.~Brown, editors, \emph{Monitoring Biodiversity}, pages 148--173. Routledge, 1st edition, 2023.
\newblock \doi{10.4324/9781003179245-8}.

\bibitem[Arias et~al.(2023)Arias, Feijoo, and Moreira]{arias2023}
A.~Arias, G.~Feijoo, and M.~T. Moreira.
\newblock How could artificial intelligence be used to increase the potential of biorefineries in the near future? a review.
\newblock \emph{Environmental Technology \& Innovation}, 32:\penalty0 103277, 2023.
\newblock \doi{10.1016/j.eti.2023.103277}.

\bibitem[Armenteras(2021)]{armenteras2021}
D.~Armenteras.
\newblock Guidelines for healthy global scientific collaborations.
\newblock \emph{Nature Ecology \& Evolution}, 5\penalty0 (9):\penalty0 1193--1194, 2021.
\newblock \doi{10.1038/s41559-021-01496-y}.

\bibitem[Barocas et~al.(2023)Barocas, Hardt, and Narayanan]{barocas2023}
S.~Barocas, M.~Hardt, and A.~Narayanan.
\newblock \emph{Fairness and machine learning: Limitations and opportunities}.
\newblock MIT Press, 2023.

\bibitem[Beery(2021)]{beery2021scaling}
S.~Beery.
\newblock Scaling biodiversity monitoring for the data age.
\newblock \emph{XRDS: Crossroads, The ACM Magazine for Students}, 27\penalty0 (4):\penalty0 14--18, 2021.

\bibitem[Beloiu et~al.(2023)Beloiu, Heinzmann, Rehush, Gessler, and Griess]{beloiu2023}
M.~Beloiu, L.~Heinzmann, N.~Rehush, A.~Gessler, and V.~C. Griess.
\newblock Individual tree-crown detection and species identification in heterogeneous forests using aerial rgb imagery and deep learning.
\newblock \emph{Remote Sensing}, 15\penalty0 (5):\penalty0 1463, 2023.
\newblock \doi{10.3390/rs15051463}.

\bibitem[Bender et~al.(2021)Bender, Gebru, McMillan-Major, and Shmitchell]{bender2021}
E.~M. Bender, T.~Gebru, A.~McMillan-Major, and S.~Shmitchell.
\newblock On the dangers of stochastic parrots: Can language models be too big?
\newblock In \emph{Proceedings of the 2021 ACM Conference on Fairness, Accountability, and Transparency}, pages 610--623, 2021.
\newblock \doi{10.1145/3442188.3445922}.

\bibitem[Bessey et~al.(2021)Bessey, Neil~Jarman, Simpson, Miller, Stewart, Kenneth~Keesing, and Berry]{bessey2021}
C.~Bessey, S.~Neil~Jarman, T.~Simpson, H.~Miller, T.~Stewart, J.~Kenneth~Keesing, and O.~Berry.
\newblock Passive edna collection enhances aquatic biodiversity analysis.
\newblock \emph{Communications Biology}, 4\penalty0 (1):\penalty0 236, 2021.
\newblock \doi{10.1038/s42003-021-01760-8}.

\bibitem[Bommasani et~al.(2022)Bommasani, Hudson, Adeli, Altman, Arora, von Arx, Bernstein, Bohg, Bosselut, Brunskill, Brynjolfsson, Buch, Card, Castellon, Chatterji, Chen, Creel, Davis, Demszky, and Liang]{bommasani2022}
R.~Bommasani, D.~A. Hudson, E.~Adeli, R.~Altman, S.~Arora, S.~von Arx, M.~S. Bernstein, J.~Bohg, A.~Bosselut, E.~Brunskill, E.~Brynjolfsson, S.~Buch, D.~Card, R.~Castellon, N.~Chatterji, A.~Chen, K.~Creel, J.~Q. Davis, D.~Demszky, and P.~Liang.
\newblock On the opportunities and risks of foundation models.
\newblock \emph{arXiv}, July 2022.
\newblock URL \url{http://arxiv.org/abs/2108.07258}.

\bibitem[Bonney et~al.(2009)Bonney, Cooper, Dickinson, Kelling, Phillips, Rosenberg, and Shirk]{bonney2009}
R.~Bonney, C.~B. Cooper, J.~Dickinson, S.~Kelling, T.~Phillips, K.~V. Rosenberg, and J.~Shirk.
\newblock Citizen science: A developing tool for expanding science knowledge and scientific literacy.
\newblock \emph{BioScience}, 59\penalty0 (11):\penalty0 977--984, 2009.
\newblock \doi{10.1525/bio.2009.59.11.9}.

\bibitem[Boyer et~al.(2016)Boyer, Mercier, Bonin, Le~Bras, Taberlet, and Coissac]{boyer2016}
F.~Boyer, C.~Mercier, A.~Bonin, Y.~Le~Bras, P.~Taberlet, and E.~Coissac.
\newblock obitools: A unix-inspired software package for dna metabarcoding.
\newblock \emph{Molecular Ecology Resources}, 16\penalty0 (1):\penalty0 176--182, 2016.
\newblock \doi{10.1111/1755-0998.12428}.

\bibitem[Burns et~al.(2023)Burns, Sankar-King, Dell’Orto, and Roma]{burns2023}
M.~Burns, S.~Sankar-King, P.~Dell’Orto, and E.~Roma.
\newblock Using ai to build stronger connections with customers.
\newblock \emph{Harvard Business Review}, August 2023.
\newblock URL \url{https://hbr.org/2023/08/using-ai-to-build-stronger-connections-with-customers}.

\bibitem[Carroll et~al.(2021)Carroll, Herczog, Hudson, Russell, and Stall]{carroll2021}
S.~R. Carroll, E.~Herczog, M.~Hudson, K.~Russell, and S.~Stall.
\newblock Operationalizing the care and fair principles for indigenous data futures.
\newblock \emph{Scientific Data}, 8\penalty0 (1):\penalty0 108, 2021.
\newblock \doi{10.1038/s41597-021-00892-0}.

\bibitem[Cañas et~al.(2023)Cañas, Toro-Gómez, Sugai, Benítez~Restrepo, Rudas, Posso~Bautista, and Ulloa]{canas2023}
J.~S. Cañas, M.~P. Toro-Gómez, L.~S.~M. Sugai, H.~D. Benítez~Restrepo, J.~Rudas, B.~Posso~Bautista, and J.~S. Ulloa.
\newblock A dataset for benchmarking neotropical anuran calls identification in passive acoustic monitoring.
\newblock \emph{Scientific Data}, 10\penalty0 (1):\penalty0 771, 2023.
\newblock \doi{10.1038/s41597-023-02666-2}.

\bibitem[Chapman et~al.(2024)Chapman, Goldstein, Schell, Brashares, Carter, Ellis-Soto, Faxon, Goldstein, Halpern, Longdon, Norman, O’Rourke, Scoville, Xu, and Boettiger]{chapman2024}
M.~Chapman, B.~R. Goldstein, C.~J. Schell, J.~S. Brashares, N.~H. Carter, D.~Ellis-Soto, H.~O. Faxon, J.~E. Goldstein, B.~S. Halpern, J.~Longdon, K.~E.~A. Norman, D.~O’Rourke, C.~Scoville, L.~Xu, and C.~Boettiger.
\newblock Biodiversity monitoring for a just planetary future.
\newblock \emph{Science}, 383\penalty0 (6678):\penalty0 34--36, 2024.
\newblock \doi{10.1126/science.adh8874}.

\bibitem[Cole et~al.(2023)Cole, Stathatos, Lütjens, Sharma, Kay, Parham, Kellenberger, and Beery]{cole2023}
E.~Cole, S.~Stathatos, B.~Lütjens, T.~Sharma, J.~Kay, J.~Parham, B.~Kellenberger, and S.~Beery.
\newblock Teaching computer vision for ecology.
\newblock \emph{arXiv}, 2023.
\newblock \doi{10.48550/ARXIV.2301.02211}.

\bibitem[Collen et~al.(2008)Collen, Ram, Zamin, and McRae]{collen2008}
B.~Collen, M.~Ram, T.~Zamin, and L.~McRae.
\newblock The tropical biodiversity data gap: Addressing disparity in global monitoring.
\newblock \emph{Tropical Conservation Science}, 1\penalty0 (2):\penalty0 75--88, 2008.
\newblock \doi{10.1177/19400829080010020}.

\bibitem[Crawford(2021)]{crawford2021}
K.~Crawford.
\newblock \emph{The atlas of AI: Power, politics, and the planetary costs of artificial intelligence}.
\newblock Yale University Press, 2021.

\bibitem[Dastin(2022)]{dastin2022}
J.~Dastin.
\newblock Amazon scraps secret ai recruiting tool that showed bias against women.
\newblock In \emph{Ethics of Data and Analytics}. Auerbach Publications, 2022.

\bibitem[de~Ambiente~y Desarrollo~Sostenible(2024)]{ministerio2024}
M.~de~Ambiente~y Desarrollo~Sostenible.
\newblock Plan de acción de biodiversidad de colombia al 2030 / colombia.
\newblock Technical report, Ministerio de Ambiente y Desarrollo Sostenible, 2024.
\newblock Redacción y edición general Laura Camila Bermúdez Wilches, María Carolina Pinilla Herrera, Lizeth Carolina Quiroga Cubillos.

\bibitem[De~Araújo(2024)]{de2024}
C.~B. De~Araújo.
\newblock We must not fool ourselves: A reply to sethi et al. on the use of birdnet to classify neotropical birdcalls.
\newblock \emph{Proceedings of the National Academy of Sciences}, 121\penalty0 (51):\penalty0 e2419635121, 2024.
\newblock \doi{10.1073/pnas.2419635121}.

\bibitem[Deiner et~al.(2017)Deiner, Bik, Mächler, Seymour, Lacoursière-Roussel, Altermatt, Creer, Bista, Lodge, De~Vere, et~al.]{deiner2017}
K.~Deiner, H.~M. Bik, E.~Mächler, M.~Seymour, A.~Lacoursière-Roussel, F.~Altermatt, S.~Creer, I.~Bista, D.~M. Lodge, N.~De~Vere, et~al.
\newblock Environmental dna metabarcoding: Transforming how we survey animal and plant communities.
\newblock \emph{Molecular Ecology}, 26\penalty0 (21):\penalty0 5872--5895, 2017.

\bibitem[DNP(2018)]{dnp2018}
DNP.
\newblock Política de crecimiento verde (conpes 3934).
\newblock Technical report, Departamento Nacional de Planeación de Colombia, 2018.

\bibitem[DNP(2025)]{dnp2025}
DNP.
\newblock Política nacional de inteligencia artificial (conpes 4144).
\newblock Technical report, Departamento Nacional de Planeación de Colombia, 2025.

\bibitem[Doshi-Velez and Kim(2017)]{doshi2017}
F.~Doshi-Velez and B.~Kim.
\newblock Towards a rigorous science of interpretable machine learning.
\newblock \emph{arXiv}, 2017.
\newblock URL \url{https://arxiv.org/abs/1702.08608}.

\bibitem[Fletcher(2014)]{fletcher2014}
N.~H. Fletcher.
\newblock Animal bioacoustics.
\newblock In T.~D. Rossing, editor, \emph{Springer Handbook of Acoustics}, pages 821--841. Springer New York, 2014.
\newblock \doi{10.1007/978-1-4939-0755-7_19}.

\bibitem[Flück et~al.(2022)Flück, Mathon, Manel, Valentini, Dejean, Albouy, Mouillot, Thuiller, Murienne, Brosse, and Pellissier]{fluck2022}
B.~Flück, L.~Mathon, S.~Manel, A.~Valentini, T.~Dejean, C.~Albouy, D.~Mouillot, W.~Thuiller, J.~Murienne, S.~Brosse, and L.~Pellissier.
\newblock Applying convolutional neural networks to speed up environmental dna annotation in a highly diverse ecosystem.
\newblock \emph{Scientific Reports}, 12\penalty0 (1):\penalty0 10247, 2022.
\newblock \doi{10.1038/s41598-022-13412-w}.

\bibitem[Galindo et~al.(2021)Galindo, Perset, and Sheeka]{galindo2021}
L.~Galindo, K.~Perset, and F.~Sheeka.
\newblock An overview of national ai strategies and policies.
\newblock \emph{OECD Going Digital Toolkit Notes}, \penalty0 (14), 2021.

\bibitem[Giraldo-Zuluaga et~al.(2019)Giraldo-Zuluaga, Salazar, Gomez, and Diaz-Pulido]{giraldo2019}
J.~H. Giraldo-Zuluaga, A.~Salazar, A.~Gomez, and A.~Diaz-Pulido.
\newblock Camera-trap images segmentation using multi-layer robust principal component analysis.
\newblock \emph{The Visual Computer}, 35:\penalty0 335--347, 2019.
\newblock \doi{10.1007/s00371-017-1463-9}.

\bibitem[Gori et~al.(2022)Gori, Ulian, Bernal, and Diazgranados]{gori2022understanding}
B.~Gori, T.~Ulian, H.~Bernal, and M.~Diazgranados.
\newblock Understanding the diversity and biogeography of colombian edible plants.
\newblock \emph{Scientific Reports}, 12\penalty0 (1):\penalty0 7835, 2022.

\bibitem[GPAI(2022)]{gpai2022}
GPAI.
\newblock Biodiversity \& artificial intelligence, opportunities and recommendations.
\newblock Technical report, Global Partnership on AI, 2022.

\bibitem[Haelewaters et~al.(2021)Haelewaters, Hofmann, and Romero-Olivares]{haelewaters2021}
D.~Haelewaters, T.~A. Hofmann, and A.~L. Romero-Olivares.
\newblock Ten simple rules for global north researchers to stop perpetuating helicopter research in the global south.
\newblock \emph{PLOS Computational Biology}, 17\penalty0 (8):\penalty0 e1009277, 2021.
\newblock \doi{10.1371/journal.pcbi.1009277}.

\bibitem[Harvey and Gericke(2011)]{harvey2011}
A.~L. Harvey and N.~Gericke.
\newblock Bioprospecting: Creating a value for biodiversity.
\newblock In \emph{Research in Biodiversity: Models and Applications}, pages 323--338. 2011.

\bibitem[Hassan et~al.(2022)Hassan, Sabreena, Ganai, Almalki, Gafur, and Sayyed]{hassan2022}
S.~Hassan, P.~Sabreena, Poczai, B.~A. Ganai, W.~H. Almalki, A.~Gafur, and R.~Z. Sayyed.
\newblock Environmental dna metabarcoding: A novel contrivance for documenting terrestrial biodiversity.
\newblock \emph{Biology}, 11\penalty0 (9):\penalty0 1297, 2022.
\newblock \doi{10.3390/biology11091297}.

\bibitem[Hastie et~al.(2009)Hastie, Tibshirani, Friedman, and Friedman]{hastie2009elements}
T.~Hastie, R.~Tibshirani, J.~H. Friedman, and J.~H. Friedman.
\newblock \emph{The elements of statistical learning: data mining, inference, and prediction}, volume~2.
\newblock Springer, 2009.

\bibitem[Hirschfeld et~al.(2023)Hirschfeld, Faria, and Fonseca]{hirschfeld2023}
M.~N.~C. Hirschfeld, L.~R.~R. Faria, and C.~R. Fonseca.
\newblock Avoid the reproduction of coloniality in decolonial studies in ecology.
\newblock \emph{Nature Ecology \& Evolution}, 7\penalty0 (3):\penalty0 306--309, 2023.
\newblock \doi{10.1038/s41559-022-01971-0}.

\bibitem[Hunter and Brehm(2003)]{hunter2003}
L.~M. Hunter and J.~Brehm.
\newblock Qualitative insight into public knowledge of, and concern with, biodiversity.
\newblock \emph{Human Ecology}, pages 309--320, 2003.

\bibitem[Jobin et~al.(2019)Jobin, Ienca, and Vayena]{jobin2019}
A.~Jobin, M.~Ienca, and E.~Vayena.
\newblock The global landscape of ai ethics guidelines.
\newblock \emph{Nature Machine Intelligence}, 1\penalty0 (9):\penalty0 389--399, 2019.
\newblock \doi{10.1038/s42256-019-0088-2}.

\bibitem[Kahl et~al.(2021)Kahl, Wood, Eibl, and Klinck]{kahl2021}
S.~Kahl, C.~M. Wood, M.~Eibl, and H.~Klinck.
\newblock Birdnet: A deep learning solution for avian diversity monitoring.
\newblock \emph{Ecological Informatics}, 61:\penalty0 101236, 2021.
\newblock \doi{10.1016/j.ecoinf.2021.101236}.

\bibitem[Kamble and Singh(2022)]{kamble2022}
A.~D. Kamble and H.~Singh.
\newblock Finding novel enzymes by in silico bioprospecting approach.
\newblock In M.~Kuddus and C.~N. Aguilar, editors, \emph{Value-Addition in Food Products and Processing Through Enzyme Technology}, pages 347--364. Academic Press, 2022.
\newblock \doi{10.1016/B978-0-323-89929-1.00028-7}.

\bibitem[Khipu(2025)]{khipu}
Khipu, 2025.
\newblock URL \url{https://khipu.ai/}.
\newblock Accessed: 2025-01-03.

\bibitem[Kimura et~al.(2022)Kimura, Yamanaka, and Nakashima]{kimura2022}
M.~Kimura, H.~Yamanaka, and Y.~Nakashima.
\newblock Application of machine learning to environmental dna metabarcoding.
\newblock \emph{IEEE Access}, 10:\penalty0 101790--101794, 2022.
\newblock \doi{10.1109/ACCESS.2022.3207173}.

\bibitem[Lamperti et~al.(2023)Lamperti, Sanchez, Si~Moussi, Mouillot, Albouy, Flück, Bruno, Valentini, Pellissier, and Manel]{lamperti2023}
L.~Lamperti, T.~Sanchez, S.~Si~Moussi, D.~Mouillot, C.~Albouy, B.~Flück, M.~Bruno, A.~Valentini, L.~Pellissier, and S.~Manel.
\newblock New deep learning‐based methods for visualizing ecosystem properties using environmental dna metabarcoding data.
\newblock \emph{Molecular Ecology Resources}, 23\penalty0 (8):\penalty0 1946--1958, 2023.
\newblock \doi{10.1111/1755-0998.13861}.

\bibitem[LeCun et~al.(2015)LeCun, Bengio, and Hinton]{lecun2015}
Y.~LeCun, Y.~Bengio, and G.~Hinton.
\newblock Deep learning.
\newblock \emph{Nature}, 521\penalty0 (7553):\penalty0 436--444, 2015.
\newblock \doi{10.1038/nature14539}.

\bibitem[Li et~al.(2022)Li, Du, and He]{li2022}
D.~Li, P.~Du, and H.~He.
\newblock Artificial intelligence-based sustainable development of smart heritage tourism.
\newblock \emph{Wireless Communications and Mobile Computing}, 2022:\penalty0 1--13, 2022.
\newblock \doi{10.1155/2022/5441170}.

\bibitem[Lucas(2020)]{lucas2020}
T.~C.~D. Lucas.
\newblock A translucent box: Interpretable machine learning in ecology.
\newblock \emph{Ecological Monographs}, 90\penalty0 (4):\penalty0 e01422, 2020.
\newblock \doi{10.1002/ecm.1422}.

\bibitem[LXAI(2025)]{latinX}
LXAI, 2025.
\newblock URL \url{https://www.latinxinai.org/}.
\newblock Accessed: 2025-01-03.

\bibitem[Ma et~al.(2024)Ma, Wei, Zhu, Zhao, Wu, Chen, Li, and Liu]{ma2024}
D.~Ma, J.~Wei, L.~Zhu, F.~Zhao, H.~Wu, X.~Chen, Y.~Li, and M.~Liu.
\newblock Semi-supervised learning advances species recognition for aquatic biodiversity monitoring.
\newblock \emph{Frontiers in Marine Science}, 11, 2024.
\newblock \doi{10.3389/fmars.2024.1373755}.

\bibitem[Mac~Aodha et~al.(2022)Mac~Aodha, Martínez~Balvanera, Damstra, Cooke, Eichinski, Browning, and Jones]{mac2022}
O.~M. Mac~Aodha, S.~Martínez~Balvanera, E.~Damstra, M.~Cooke, P.~Eichinski, E.~Browning, and K.~E. Jones.
\newblock Towards a general approach for bat echolocation detection and classification.
\newblock \emph{bioRxiv}, 2022.
\newblock \doi{10.1101/2022.12.14.520490}.

\bibitem[Maldonado(2020)]{maldonado2020}
N.~F.~J. Maldonado.
\newblock Libre comercio y ventajas competitivas en colombia: Mercado y estado. una visión desde la teoría de ventajas competitivas de michael e. porter.
\newblock \emph{Divergencia}, 27:\penalty0 7--18, 2020.

\bibitem[Mendoza-Henao et~al.(2023)Mendoza-Henao, Acevedo-Charry, Mart{\'\i}nez-Medina, Barona-Cort{\'e}s, C{\'o}rdoba-C{\'o}rdoba, Caycedo-Rosales, Ulloa, Borja-Acosta, Buitrago-Cardona, and Pantoja-S{\'a}nchez]{mendoza2023past}
A.~M. Mendoza-Henao, O.~Acevedo-Charry, D.~Mart{\'\i}nez-Medina, E.~Barona-Cort{\'e}s, S.~C{\'o}rdoba-C{\'o}rdoba, P.~Caycedo-Rosales, J.~S. Ulloa, K.~G. Borja-Acosta, A.~Buitrago-Cardona, and H.~Pantoja-S{\'a}nchez.
\newblock Past, present, and future of a tropical sounds collection from colombia.
\newblock \emph{Bioacoustics}, 32\penalty0 (4):\penalty0 474--490, 2023.

\bibitem[Messeri and Crockett(2024)]{messeri2024artificial}
L.~Messeri and M.~Crockett.
\newblock Artificial intelligence and illusions of understanding in scientific research.
\newblock \emph{Nature}, 627\penalty0 (8002):\penalty0 49--58, 2024.

\bibitem[Miao et~al.(2023)Miao, Elizalde, Deshmukh, Kitzes, Wang, Dodhia, and Ferres]{miao2023}
Z.~Miao, B.~Elizalde, S.~Deshmukh, J.~Kitzes, H.~Wang, R.~Dodhia, and J.~M.~L. Ferres.
\newblock Zero-shot transfer for wildlife bioacoustics detection.
\newblock 2023.
\newblock \doi{10.21203/rs.3.rs-3180218/v1}.

\bibitem[Miller(2022)]{miller2022}
G.~J. Miller.
\newblock Stakeholder roles in artificial intelligence projects.
\newblock \emph{Project Leadership and Society}, 3:\penalty0 100068, 2022.
\newblock \doi{10.1016/j.plas.2022.100068}.

\bibitem[Mohamed et~al.(2020)Mohamed, Png, and Isaac]{mohamed2020}
S.~Mohamed, M.-T. Png, and W.~Isaac.
\newblock Decolonial ai: Decolonial theory as sociotechnical foresight in artificial intelligence.
\newblock \emph{Philosophy \& Technology}, 33\penalty0 (4):\penalty0 659--684, 2020.
\newblock \doi{10.1007/s13347-020-00405-8}.

\bibitem[Mora et~al.(2011)Mora, Tittensor, Adl, Simpson, and Worm]{mora2011}
C.~Mora, D.~P. Tittensor, S.~Adl, A.~G.~B. Simpson, and B.~Worm.
\newblock How many species are there on earth and in the ocean?
\newblock \emph{PLoS Biology}, 9\penalty0 (8):\penalty0 e1001127, 2011.
\newblock \doi{10.1371/journal.pbio.1001127}.

\bibitem[Muha et~al.(2017)Muha, Rodríguez-Rey, Rolla, and Tricarico]{muha2017}
T.~P. Muha, M.~Rodríguez-Rey, M.~Rolla, and E.~Tricarico.
\newblock Using environmental dna to improve species distribution models for freshwater invaders.
\newblock \emph{Frontiers in Ecology and Evolution}, 5:\penalty0 158, 2017.
\newblock \doi{10.3389/fevo.2017.00158}.

\bibitem[Muscarella et~al.(2014)Muscarella, Galante, Soley‐Guardia, Boria, Kass, Uriarte, and Anderson]{muscarella2014}
R.~Muscarella, P.~J. Galante, M.~Soley‐Guardia, R.~A. Boria, J.~M. Kass, M.~Uriarte, and R.~P. Anderson.
\newblock Enm eval: An r package for conducting spatially independent evaluations and estimating optimal model complexity for maxent ecological niche models.
\newblock \emph{Methods in Ecology and Evolution}, 5\penalty0 (11):\penalty0 1198--1205, 2014.
\newblock \doi{10.1111/2041-210X.12261}.

\bibitem[Nehoshtan et~al.(2021)Nehoshtan, Carmon, Yaniv, Ayal, and Rotem]{nehoshtan2021}
Y.~Nehoshtan, E.~Carmon, O.~Yaniv, S.~Ayal, and O.~Rotem.
\newblock Robust seed germination prediction using deep learning and rgb image data.
\newblock \emph{Scientific Reports}, 11:\penalty0 22030, 2021.
\newblock \doi{10.1038/s41598-021-01712-6}.

\bibitem[Nicolson and Tucker(2017)]{nicolson2017}
N.~Nicolson and A.~Tucker.
\newblock Identifying novel features from specimen data for the prediction of valuable collection trips.
\newblock In N.~Adams, A.~Tucker, and D.~Weston, editors, \emph{Advances in Intelligent Data Analysis XVI}, pages 235--246. Springer International Publishing, 2017.
\newblock \doi{10.1007/978-3-319-68765-0_20}.

\bibitem[Nussinov et~al.(2022)Nussinov, Zhang, Liu, and Jang]{nussinov2022}
R.~Nussinov, M.~Zhang, Y.~Liu, and H.~Jang.
\newblock Alphafold, artificial intelligence (ai), and allostery.
\newblock \emph{The Journal of Physical Chemistry B}, 126\penalty0 (34):\penalty0 6372--6383, 2022.
\newblock \doi{10.1021/acs.jpcb.2c04346}.

\bibitem[Oliver et~al.(2023{\natexlab{a}})Oliver, Flemons, and Michael]{oliver2023b}
J.~Oliver, P.~Flemons, and K.~Michael.
\newblock Advancing the productivity of science with citizen science and artificial intelligence.
\newblock In \emph{Artificial Intelligence in Science: Challenges, Opportunities and the Future of Research}, page 148. OECD Publishing, 2023{\natexlab{a}}.

\bibitem[Oliver et~al.(2023{\natexlab{b}})Oliver, Iannarilli, Ahumada, Fegraus, Flores, Kays, and Jetz]{oliver2023a}
R.~Y. Oliver, F.~Iannarilli, J.~Ahumada, E.~Fegraus, N.~Flores, R.~Kays, and W.~Jetz.
\newblock Camera trapping expands the view into global biodiversity and its change.
\newblock \emph{Philosophical Transactions of the Royal Society B}, 378\penalty0 (1881):\penalty0 20220232, 2023{\natexlab{b}}.

\bibitem[Paic(2021)]{paic2021}
A.~Paic.
\newblock Open science—enabling discovery in the digital age.
\newblock Technical report, OECD, 2021.

\bibitem[para América Latina y~el Caribe~(CEPAL)(2022)]{comision22}
C.~E. para América Latina y~el Caribe~(CEPAL).
\newblock Acuerdo regional sobre el acceso a la información, la participación pública y el acceso a la justicia en asuntos ambientales en américa latina y el caribe.
\newblock \emph{Santiago}, 2022.
\newblock \doi{LC/PUB.2018/8/Rev.1}.

\bibitem[Pascual et~al.(2023)Pascual, Balvanera, Anderson, Chaplin-Kramer, Christie, González-Jiménez, Martin, Raymond, Termansen, Vatn, Athayde, Baptiste, Barton, Jacobs, Kelemen, Kumar, Lazos, Mwampamba, Nakangu, and Zent]{pascual2023}
U.~Pascual, P.~Balvanera, C.~B. Anderson, R.~Chaplin-Kramer, M.~Christie, D.~González-Jiménez, A.~Martin, C.~M. Raymond, M.~Termansen, A.~Vatn, S.~Athayde, B.~Baptiste, D.~N. Barton, S.~Jacobs, E.~Kelemen, R.~Kumar, E.~Lazos, T.~H. Mwampamba, B.~Nakangu, and E.~Zent.
\newblock Diverse values of nature for sustainability.
\newblock \emph{Nature}, 620\penalty0 (7975):\penalty0 813--823, 2023.
\newblock \doi{10.1038/s41586-023-06406-9}.

\bibitem[Phillips and Dudík(2008)]{phillips2008}
S.~J. Phillips and M.~Dudík.
\newblock Modeling of species distributions with maxent: New extensions and a comprehensive evaluation.
\newblock \emph{Ecography}, 31\penalty0 (2):\penalty0 161--175, 2008.
\newblock \doi{10.1111/j.0906-7590.2008.5203.x}.

\bibitem[Pulido et~al.(2018)Pulido, Isaza, and Diaz-Pulido]{pulido2018}
L.~F. Pulido, C.~Isaza, and D.~P. Diaz-Pulido.
\newblock Naira iii.
\newblock \emph{Mammalogy Notes}, 5\penalty0 (1-2):\penalty0 39--44, 2018.
\newblock \doi{10.47603/manovol5n1.39-44}.

\bibitem[Rahim et~al.(2023)Rahim, Bodnar, Qasim, Jawad, and Ahmed]{rahim2023}
F.~Rahim, N.~Bodnar, N.~H. Qasim, A.~M. Jawad, and O.~S. Ahmed.
\newblock Integrating machine learning in environmental dna metabarcoding for improved biodiversity assessment: A review and analysis of recent studies.
\newblock 2023.
\newblock \doi{10.21203/rs.3.rs-2823060/v1}.

\bibitem[Reynolds et~al.(2024)Reynolds, Beery, Burgess, Burgman, Butchart, Cooke, and Sutherland]{reynolds2024}
S.~A. Reynolds, S.~Beery, N.~Burgess, M.~Burgman, S.~H. Butchart, S.~J. Cooke, and W.~J. Sutherland.
\newblock The potential for ai to revolutionize conservation: a horizon scan.
\newblock \emph{Trends in Ecology \& Evolution}, 2024.
\newblock \doi{10.1016/j.tree.2024.11.013}.

\bibitem[Richardson et~al.(2023)Richardson, Steffen, Lucht, Bendtsen, Cornell, Donges, Drüke, Fetzer, Bala, Von~Bloh, Feulner, Fiedler, Gerten, Gleeson, Hofmann, Huiskamp, Kummu, Mohan, Nogués-Bravo, and Rockström]{richardson2023}
K.~Richardson, W.~Steffen, W.~Lucht, J.~Bendtsen, S.~E. Cornell, J.~F. Donges, M.~Drüke, I.~Fetzer, G.~Bala, W.~Von~Bloh, G.~Feulner, S.~Fiedler, D.~Gerten, T.~Gleeson, M.~Hofmann, W.~Huiskamp, M.~Kummu, C.~Mohan, D.~Nogués-Bravo, and J.~Rockström.
\newblock Earth beyond six of nine planetary boundaries.
\newblock \emph{Science Advances}, 9\penalty0 (37):\penalty0 eadh2458, 2023.
\newblock \doi{10.1126/sciadv.adh2458}.

\bibitem[Rivera~Cusicanqui(2012)]{rivera2012}
S.~Rivera~Cusicanqui.
\newblock Ch'ixinakax utxiwa: A reflection on the practices and discourses of decolonization.
\newblock \emph{South Atlantic Quarterly}, 111\penalty0 (1):\penalty0 95--109, 2012.

\bibitem[Robles-Fernandez et~al.(2022)Robles-Fernandez, Santiago-Alarcon, and Lira-Noriega]{robles2022}
A.~L. Robles-Fernandez, D.~Santiago-Alarcon, and A.~Lira-Noriega.
\newblock Wildlife susceptibility to infectious diseases at global scales.
\newblock \emph{Proceedings of the National Academy of Sciences}, 119\penalty0 (35):\penalty0 e2122851119, 2022.
\newblock \doi{10.1073/pnas.2122851119}.

\bibitem[Rohde et~al.(2024)Rohde, Wagner, Meyer, Reinhard, Voss, Petschow, and Mollen]{rohde2024}
F.~Rohde, J.~Wagner, A.~Meyer, P.~Reinhard, M.~Voss, U.~Petschow, and A.~Mollen.
\newblock Broadening the perspective for sustainable artificial intelligence: sustainability criteria and indicators for artificial intelligence systems.
\newblock \emph{Current Opinion in Environmental Sustainability}, 66:\penalty0 101411, 2024.
\newblock \doi{10.1016/j.cosust.2023.101411}.

\bibitem[Rolnick et~al.(2023)Rolnick, Donti, Kaack, Kochanski, Lacoste, Sankaran, Ross, Milojevic-Dupont, Jaques, Waldman-Brown, Luccioni, Maharaj, Sherwin, Mukkavilli, Kording, Gomes, Ng, Hassabis, Platt, and Bengio]{rolnick2023}
D.~Rolnick, P.~L. Donti, L.~H. Kaack, K.~Kochanski, A.~Lacoste, K.~Sankaran, A.~S. Ross, N.~Milojevic-Dupont, N.~Jaques, A.~Waldman-Brown, A.~S. Luccioni, T.~Maharaj, E.~D. Sherwin, S.~K. Mukkavilli, K.~P. Kording, C.~P. Gomes, A.~Y. Ng, D.~Hassabis, J.~C. Platt, and Y.~Bengio.
\newblock Tackling climate change with machine learning.
\newblock \emph{ACM Computing Surveys}, 55\penalty0 (2):\penalty0 1--96, 2023.
\newblock \doi{10.1145/3485128}.

\bibitem[Rudin(2019)]{rudin2019}
C.~Rudin.
\newblock Stop explaining black box machine learning models for high stakes decisions and use interpretable models instead.
\newblock \emph{Nature machine intelligence}, 1\penalty0 (5):\penalty0 206--215, 2019.
\newblock \doi{10.1038/s42256-019-0048-x}.

\bibitem[Ruiz~Puentes et~al.(2022)Ruiz~Puentes, Henao, Cifuentes, Muñoz-Camargo, Reyes, Cruz, and Arbeláez]{ruiz2022}
P.~Ruiz~Puentes, M.~C. Henao, J.~Cifuentes, C.~Muñoz-Camargo, L.~H. Reyes, J.~C. Cruz, and P.~Arbeláez.
\newblock Rational discovery of antimicrobial peptides by means of artificial intelligence.
\newblock \emph{Membranes}, 12\penalty0 (7):\penalty0 708, 2022.
\newblock \doi{10.3390/membranes12070708}.

\bibitem[Russell and Norvig(2016)]{russell2016}
S.~J. Russell and P.~Norvig.
\newblock \emph{Artificial intelligence: A modern approach}.
\newblock Pearson, 2016.

\bibitem[Ryan and Stahl(2021)]{ryan2021}
M.~Ryan and B.~C. Stahl.
\newblock Artificial intelligence ethics guidelines for developers and users: Clarifying their content and normative implications.
\newblock \emph{Journal of Information, Communication and Ethics in Society}, 19\penalty0 (1):\penalty0 61--86, 2021.
\newblock \doi{10.1108/JICES-12-2019-0138}.

\bibitem[Salili-James et~al.(2023)Salili-James, Scott, Livermore, Price, Dupont, Hardy, and Smith]{salili2023}
A.~Salili-James, B.~Scott, L.~Livermore, B.~Price, S.~Dupont, H.~Hardy, and V.~Smith.
\newblock Ai-accelerated digitisation of insect collections: The next generation of angled label image capture equipment (alice).
\newblock \emph{Biodiversity Information Science and Standards}, 7:\penalty0 e112742, 2023.
\newblock \doi{10.3897/biss.7.112742}.

\bibitem[Sanchez-Pi et~al.(2021)Sanchez-Pi, Martí, Garcia, Yates, Vellasco, and Coello]{sanchez2021}
N.~Sanchez-Pi, L.~Martí, A.~C.~B. Garcia, R.~B. Yates, M.~Vellasco, and C.~A.~C. Coello.
\newblock A roadmap for ai in latin america.
\newblock In \emph{Side event AI in Latin America of the Global Partnership for AI (GPAI) Paris Summit}, November 2021.
\newblock URL \url{https://hal.archives-ouvertes.fr/hal-03526055}.

\bibitem[Santana et~al.(2021)Santana, do~Nascimento, Lima~e Lima, Damasceno, Nahum, Braga, and Lameira]{santana2021}
K.~Santana, L.~D. do~Nascimento, A.~Lima~e Lima, V.~Damasceno, C.~Nahum, R.~C. Braga, and J.~Lameira.
\newblock Applications of virtual screening in bioprospecting: Facts, shifts, and perspectives to explore the chemo-structural diversity of natural products.
\newblock \emph{Frontiers in Chemistry}, 9, 2021.
\newblock \doi{10.3389/fchem.2021.662688}.

\bibitem[Schermer et~al.(2018)Schermer, Hogeweg, and Caspers]{schermer2018}
M.~Schermer, L.~Hogeweg, and M.~Caspers.
\newblock Using deep learning in collection management to reduce the taxonomist’s workload.
\newblock \emph{Biodiversity Information Science and Standards}, 2:\penalty0 e25917, 2018.
\newblock \doi{10.3897/biss.2.25917}.

\bibitem[Spalding et~al.(2023)Spalding, Longley-Wood, McNulty, Constantine, Acosta-Morel, Anthony, Cole, Hall, Nickel, Schill, Schuhmann, and Tanner]{spalding2023}
M.~D. Spalding, K.~Longley-Wood, V.~P. McNulty, S.~Constantine, M.~Acosta-Morel, V.~Anthony, A.~D. Cole, G.~Hall, B.~A. Nickel, S.~R. Schill, P.~W. Schuhmann, and D.~Tanner.
\newblock Nature dependent tourism – combining big data and local knowledge.
\newblock \emph{Journal of Environmental Management}, 337:\penalty0 117696, 2023.
\newblock \doi{10.1016/j.jenvman.2023.117696}.

\bibitem[Stenhouse et~al.(2023)Stenhouse, Fisher, Lepschi, Schmidt-Lebuhn, Rodriguez, Turco, Toms, Reeson, Paris, and Thrall]{stenhouse2023}
A.~Stenhouse, N.~Fisher, B.~Lepschi, A.~Schmidt-Lebuhn, J.~Rodriguez, F.~Turco, E.~Toms, A.~Reeson, C.~Paris, and P.~Thrall.
\newblock Improving biological collections data through human-ai collaboration.
\newblock \emph{Biodiversity Information Science and Standards}, 7:\penalty0 e112488, 2023.
\newblock \doi{10.3897/biss.7.112488}.

\bibitem[Stowell(2022)]{stowell2022}
D.~Stowell.
\newblock Computational bioacoustics with deep learning: a review and roadmap.
\newblock \emph{PeerJ}, 10:\penalty0 e13152, 2022.
\newblock \doi{10.7717/peerj.13152}.

\bibitem[Sugai et~al.(2019)Sugai, Silva, Ribeiro, and Llusia]{sugai2019}
L.~S.~M. Sugai, T.~S.~F. Silva, J.~W. Ribeiro, and D.~Llusia.
\newblock Terrestrial passive acoustic monitoring: Review and perspectives.
\newblock \emph{BioScience}, 69\penalty0 (1):\penalty0 15--25, 2019.
\newblock \doi{10.1093/biosci/biy147}.

\bibitem[Södergård et~al.(2021)Södergård, Mildorf, Habyarimana, Berre, Fernandes, and Zinke-Wehlmann]{sodergard2021}
C.~Södergård, T.~Mildorf, E.~Habyarimana, A.~J. Berre, J.~A. Fernandes, and C.~Zinke-Wehlmann, editors.
\newblock \emph{Big Data in Bioeconomy: Results from the European DataBio Project}.
\newblock Springer International Publishing, 2021.
\newblock \doi{10.1007/978-3-030-71069-9}.

\bibitem[Taberlet et~al.(2028)Taberlet, Bonin, Zinger, and Coissac]{taberlet2018}
P.~Taberlet, A.~Bonin, L.~Zinger, and E.~Coissac.
\newblock \emph{Environmental DNA: For biodiversity research and monitoring}.
\newblock Oxford University Press, 2028.

\bibitem[Tambe et~al.(2022)Tambe, Fang, and Wilder]{tambe2022}
M.~Tambe, F.~Fang, and B.~Wilder, editors.
\newblock \emph{AI for Social Impact}.
\newblock 2022.
\newblock \url{https://ai4sibook.org/}.

\bibitem[Trisos et~al.(2021)Trisos, Auerbach, and Katti]{trisos2021}
C.~H. Trisos, J.~Auerbach, and M.~Katti.
\newblock Decoloniality and anti-oppressive practices for a more ethical ecology.
\newblock \emph{Nature Ecology \& Evolution}, 5\penalty0 (9):\penalty0 1205--1212, 2021.
\newblock \doi{10.1038/s41559-021-01460-w}.

\bibitem[Tuia et~al.(2022)Tuia, Kellenberger, Beery, Costelloe, Zuffi, Risse, Mathis, Mathis, Van~Langevelde, Burghardt, Kays, Klinck, Wikelski, Couzin, Van~Horn, Crofoot, Stewart, and Berger-Wolf]{tuia2022}
D.~Tuia, B.~Kellenberger, S.~Beery, B.~R. Costelloe, S.~Zuffi, B.~Risse, A.~Mathis, M.~W. Mathis, F.~Van~Langevelde, T.~Burghardt, R.~Kays, H.~Klinck, M.~Wikelski, I.~D. Couzin, G.~Van~Horn, M.~C. Crofoot, C.~V. Stewart, and T.~Berger-Wolf.
\newblock Perspectives in machine learning for wildlife conservation.
\newblock \emph{Nature Communications}, 13\penalty0 (1):\penalty0 792, 2022.
\newblock \doi{10.1038/s41467-022-27980-y}.

\bibitem[Unidas(2007)]{naciones2007}
N.~Unidas.
\newblock Capítulo 2: Convention on biological diversity.
\newblock Technical report, 2007.
\newblock URL \url{https://www.cbd.int/gbo1/chap-02.shtml}.

\bibitem[Van~Horn and Perona(2017)]{van2017}
G.~Van~Horn and P.~Perona.
\newblock The devil is in the tails: Fine-grained classification in the wild.
\newblock \emph{arXiv Preprint arXiv:1709.01450}, 2017.

\bibitem[Varoquaux et~al.(2024)Varoquaux, Luccioni, and Whittaker]{varoquaux2024}
G.~Varoquaux, A.~S. Luccioni, and M.~Whittaker.
\newblock Hype, sustainability, and the price of the bigger-is-better paradigm in ai.
\newblock \emph{arXiv preprint arXiv:2409.14160}, 2024.

\bibitem[Velásquez-Tibatá et~al.(2019)Velásquez-Tibatá, Olaya-Rodríguez, López-Lozano, Gutiérrez, González, and Londoño-Murcia]{velasquez2019}
J.~Velásquez-Tibatá, M.~H. Olaya-Rodríguez, D.~López-Lozano, C.~Gutiérrez, I.~González, and M.~C. Londoño-Murcia.
\newblock Biomodelos: A collaborative online system to map species distributions.
\newblock \emph{PLOS ONE}, 14\penalty0 (3):\penalty0 e0214522, 2019.
\newblock \doi{10.1371/journal.pone.0214522}.

\bibitem[Vicente-Saez and Martinez-Fuentes(2018)]{vicente2018}
R.~Vicente-Saez and C.~Martinez-Fuentes.
\newblock Open science now: A systematic literature review for an integrated definition.
\newblock \emph{Journal of Business Research}, 88:\penalty0 428--436, 2018.
\newblock \doi{10.1016/j.jbusres.2017.12.043}.

\bibitem[Waller(2020)]{waller2020}
J.~Waller.
\newblock Outlier detection at gbif using dbscan.
\newblock \emph{Biodiversity Information Science and Standards}, 4:\penalty0 e59412, 2020.
\newblock \doi{10.3897/biss.4.59412}.

\bibitem[Wearn et~al.(2019)Wearn, Freeman, and Jacoby]{wearn2019}
O.~R. Wearn, R.~Freeman, and D.~M.~P. Jacoby.
\newblock Responsible ai for conservation.
\newblock \emph{Nature Machine Intelligence}, 1\penalty0 (2):\penalty0 72--73, 2019.
\newblock \doi{10.1038/s42256-019-0022-7}.

\bibitem[Wilkinson et~al.(2016)Wilkinson, Dumontier, Aalbersberg, Appleton, Axton, Baak, Blomberg, Boiten, da~Silva~Santos, Bourne, Bouwman, Brookes, Clark, Crosas, Dillo, Dumon, Edmunds, Evelo, Finkers, and Mons]{wilkinson2016}
M.~D. Wilkinson, M.~Dumontier, I.~J. Aalbersberg, G.~Appleton, M.~Axton, A.~Baak, N.~Blomberg, J.-W. Boiten, L.~B. da~Silva~Santos, P.~E. Bourne, J.~Bouwman, A.~J. Brookes, T.~Clark, M.~Crosas, I.~Dillo, O.~Dumon, S.~Edmunds, C.~T. Evelo, R.~Finkers, and B.~Mons.
\newblock The fair guiding principles for scientific data management and stewardship.
\newblock \emph{Scientific Data}, 3\penalty0 (1):\penalty0 160018, 2016.
\newblock \doi{10.1038/sdata.2016.18}.

\bibitem[Wood and Kahl(2024)]{wood2024}
C.~M. Wood and S.~Kahl.
\newblock Guidelines for appropriate use of birdnet scores and other detector outputs.
\newblock \emph{Journal of Ornithology}, 165\penalty0 (3):\penalty0 777--782, 2024.
\newblock \doi{10.1007/s10336-024-02144-5}.

\bibitem[Wood et~al.(2022)Wood, Kahl, Rahaman, and Klinck]{wood2022}
C.~M. Wood, S.~Kahl, A.~Rahaman, and H.~Klinck.
\newblock The machine learning–powered birdnet app reduces barriers to global bird research by enabling citizen science participation.
\newblock \emph{PLOS Biology}, 20\penalty0 (6):\penalty0 e3001670, 2022.
\newblock \doi{10.1371/journal.pbio.3001670}.

\bibitem[Xu et~al.(2023)Xu, Rolf, Beery, Bennett, Berger-Wolf, Birch, Bondi-Kelly, Brashares, Chapman, Corso, Davies, Garg, Gaylard, Heilmayr, Kerner, Klemmer, Kumar, Mackey, Monteleoni, and Tambe]{xu2023}
L.~Xu, E.~Rolf, S.~Beery, J.~R. Bennett, T.~Berger-Wolf, T.~Birch, E.~Bondi-Kelly, J.~Brashares, M.~Chapman, A.~Corso, A.~Davies, N.~Garg, A.~Gaylard, R.~Heilmayr, H.~Kerner, K.~Klemmer, V.~Kumar, L.~Mackey, C.~Monteleoni, and M.~Tambe.
\newblock Reflections from the workshop on ai-assisted decision making for conservation.
\newblock \emph{arXiv}, July 2023.
\newblock URL \url{http://arxiv.org/abs/2307.08774}.

\bibitem[Yang et~al.(2024)Yang, Li, Lo, Zhang, Chen, Gao, U, Dai, Nakaoka, Yang, and Cheng]{yang2024}
J.~Yang, C.~Li, L.~S.~H. Lo, X.~Zhang, Z.~Chen, J.~Gao, C.~U, Z.~Dai, M.~Nakaoka, H.~Yang, and J.~Cheng.
\newblock Artificial intelligence-assisted environmental dna metabarcoding and high-resolution underwater optical imaging for noninvasive and innovative marine environmental monitoring.
\newblock \emph{Journal of Marine Science and Engineering}, 12\penalty0 (10):\penalty0 1729, 2024.
\newblock \doi{10.3390/jmse12101729}.

\bibitem[Zambrano and Sanchez-Torres(2021)]{zambrano2021}
R.~Zambrano and J.~M. Sanchez-Torres.
\newblock Ai public policies in latin america: Disruption or more of the same?
\newblock In \emph{Proceedings of the 14th International Conference on Theory and Practice of Electronic Governance}, pages 25--33, 2021.
\newblock \doi{10.1145/3494193.3494294}.

\end{thebibliography}

\clearpage
\beginsupplement

\section{Glosario de términos (Glossary of terms)}
\label{sec:supplementary}
\begin{paracol}{2}

\columnratio{0.515}

\hyphenpenalty=1000   
\tolerance=2000  

Con el fin de dar una introducción a términos técnicos relacionados con la IA y facilitar la lectura del documento, proporcionamos un breve repaso a las definiciones básicas de esta disciplina.
\begin{itemize}
    \item Inteligencia Artificial (IA): capacidad de una máquina para realizar razonamientos semánticos de alto nivel, infiriendo información a partir de datos existentes.

    \item Aprendizaje automático: tipo de IA donde las máquinas aprenden de los datos sin estar explícitamente programadas. Se utiliza en la identificación de patrones en grandes bases de datos (genéticos, ubicación de especies, climáticos, audio, imágen, etc).

    \item Aprendizaje supervisado: tipo de aprendizaje automático donde se entrena un modelo con datos etiquetados (con respuestas conocidas) para que aprenda a predecir la etiqueta de nuevos datos.

    \item Aprendizaje no supervisado: tipo de aprendizaje automático donde se entrena un modelo con datos no etiquetados para que encuentre patrones y estructuras subyacentes en los datos.

    \item Aprendizaje semi-supervisado: tipo de aprendizaje automático que combina datos etiquetados y no etiquetados para mejorar el entrenamiento del modelo, especialmente útil para identificar clases raras.

    \item Redes Neuronales Artificiales: modelos \\ matemáticos compuestos por capas de nodos interconectados que procesan datos mediante operaciones lineales y funciones de activación. Son utilizadas para aproximar funciones complejas y resolver problemas como clasificación, regresión y reconocimiento de patrones en grandes conjuntos de datos.

    \item Aprendizaje profundo: estructura especial de redes neuronales artificiales que, usando grandes cantidades de datos, permite la extracción de información de distintas fuentes de datos como texto, imágenes o audio entre otras.

    \item Redes Neuronales Convolucionales: tipo de red neuronal artificial diseñada para procesar datos generalmente bidimensionales, como imágenes. Utilizan capas en las que se aplican operaciones para extraer características que detectan patrones locales, y se emplean principalmente en tareas como reconocimiento de imágenes.

    \item Visión por computadora: rama de la IA que se centra en la interacción entre las computadoras e imágenes, permitiendo a las máquinas extraer información de imágenes y videos.

    \item Procesamiento de Lenguaje Natural (NLP): rama de la IA que se centra en la interacción entre las computadoras y el lenguaje humano, permitiendo a las máquinas comprenderlo y procesarlo.

    \item Big data: contenido de información lo suficientemente grande como para exceder el procesamiento en hardware convencional.

    \item Sistema de IA: Sistemas en los que las reglas no son definidas directamente por humanos durante la programación del algoritmo, sino generadas a través de aprendizaje automático. Estos sistemas incluyen tanto los modelos computacionales, que incorporan ciertos valores, como los conjuntos de datos utilizados para su entrenamiento \citep{rohde2024}.

\end{itemize}

\switchcolumn

To provide an introduction to technical terms related to AI and facilitate the reading of the document, we offer a brief review of the basic definitions of this discipline.

\begin{itemize}
    \item Artificial Intelligence (AI): The ability of a machine to perform high-level semantic reasoning, inferring information from existing data.

    \item Machine Learning: A type of AI where machines learn from data without being explicitly programmed. It is used to identify patterns in large databases (genetic, species location, climatic, audio, image, etc.).

    \item Supervised Learning: A type of machine learning where a model is trained with labeled data (with known answers) to learn to predict the label of new data.
    \\
    \item Unsupervised Learning: A type of machine learning where a model is trained with unlabeled data to find patterns and hidden structures.
    \\
    
    \item Semi-supervised Learning: A type of machine learning that combines labeled and unlabeled data to improve model training, especially useful for identifying rare classes.

    \item Artificial Neural Networks: Mathematical models composed of layers of interconnected nodes that process data through linear operations and activation functions. They are used to approximate complex functions and solve problems such as classification, regression, and pattern recognition in large datasets.

    \item Deep Learning: A specialized structure of artificial neural networks that, using large amounts of data, allows for information extraction from various data sources such as text, images, or audio, among others.
    
    \item Convolutional Neural Networks (CNN): A type of artificial neural network designed to process generally two-dimensional data, such as images. They use layers where operations are applied to extract features that detect local patterns and are primarily employed in tasks like image recognition.

    \item Computer Vision: branch of AI that focuses on the interaction between computers and images, enabling the extraction of information from images and videos.

    \item Natural Language Processing (NLP): A branch of AI that focuses on the interaction between computers and human language, enabling machines to understand and process it.

    \item Big Data: Information content that is large enough to exceed processing in conventional hardware.

    \item AI System: Systems in which the rules are not directly defined by humans during the algorithm's programming, but are generated through machine learning. These systems include both computational models, which incorporate certain values, and the datasets used for their training \citep{rohde2024}.

\end{itemize}

\end{paracol}

\newpage

\section{Figure \ref{fig:main_figure} translated}

\newpage

\section{Figure \ref{fig:Ddata} translated}

\begin{figure}[h]
    \centering
    \includegraphics[width=1\textwidth]{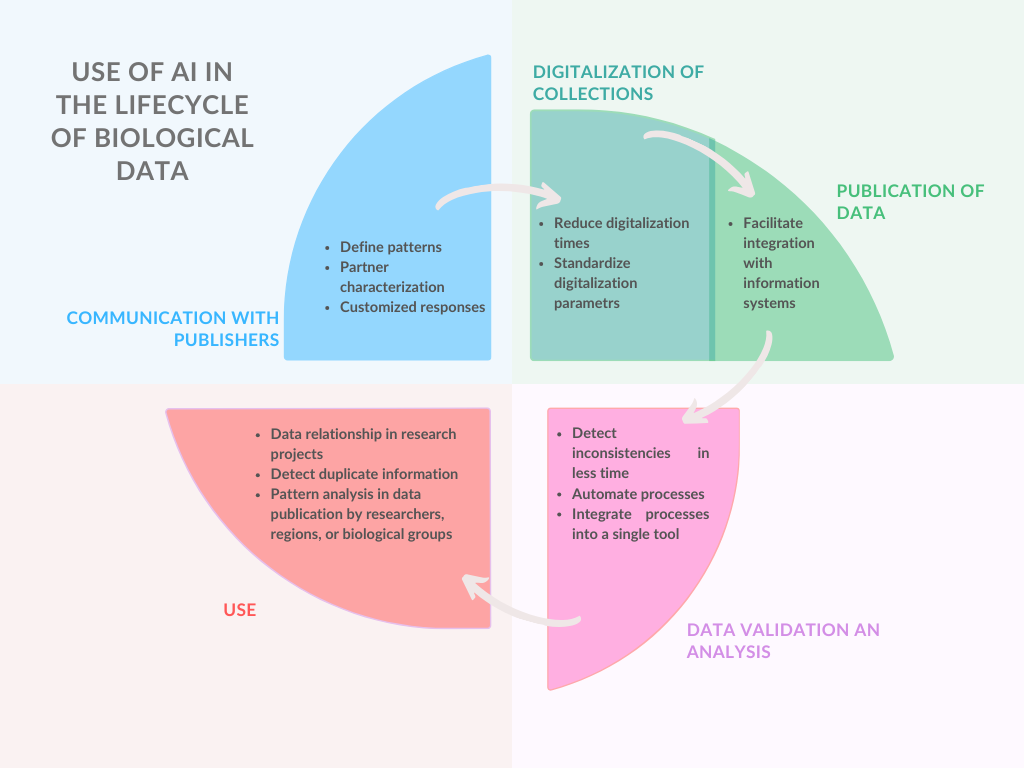}
    \caption{Representation of the flow of biological data and the role of AI at each stage. Translation of Figure \ref{fig:Ddata}.}
    \label{fig:data_eng}
\end{figure}

\newpage

\section{Table \ref{tab:tabla_esp} translated}

\begin{table}[h]
\begin{tabular}{|m{8em}|m{34em}|}
\hline
\textbf{Strategic line} &
  \textbf{Example of  IA application} \\ \hline
\multirow{2}{*}{\begin{tabular}[c]{@{}l@{}}Natural  \\
ingredients \end{tabular}} &
  \begin{tabular}[c]{@{}l@{}} 
  Discovery of bioactive compounds, which may or may not be associated with \\ the structural prediction of the proteins that compose them. This can occur at \\ the molecular, protein, or genomic level \citep{ruiz2022}, \\ \citep{nussinov2022}.
  \end{tabular} \\ \cline{2-2} 
 &
  Modeling and optimizing biorefinery production models (for obtaining bioproducts associated with natural ingredients, for example) since the large number of variables in the cascading scheme complicates the process \citep{arias2023}. \\ \hline
\multirow{2}{*}{\begin{tabular}[c]{@{}l@{}} Nature-Based  \\ Scientific Tourism \end{tabular}} &
 Modeling, measuring, and mapping the value of nature in tourism through the quantification of the spatial footprint or planning conservation and management actions in the long term (e.g., carrying capacity of the area) \citep{spalding2023}. \\ \cline{2-2} 
 &
  Application in the tourist experience (e.g., activities and tourist attractions), virtual assistance, travel planning and execution, industrial development, administrative management, among many others \citep{li2022}. \\ \hline
\multirow{2}{*}{\begin{tabular}[c]{@{}l@{}}Ecosystem \\ Restoration\end{tabular}} &
  Automatic identification and mapping of species and seed sources with restoration potential in heterogeneous forests at a large scale (i.e., a basis for inventories and specific management actions) \citep{beloiu2023}. \\ \cline{2-2} 
 &
Prediction of seed germination based on digital color images (RGB), as well as monitoring contamination, physical uniformity, and purity of the seeds \citep{nehoshtan2021}. \\ \hline
\end{tabular}
\caption{Examples of AI applications in the context of sustainable use and exploitation of biodiversity and its ecosystem services (ESS) in four strategic lines for Colombia. Translation of Table \ref{tab:tabla_esp}.
}
\label{tab:tabla_eng}
\end{table}

\end{document}